\documentclass[useAMS,usenatbib,referee]{mn2e}

\usepackage{color}
\usepackage{graphicx}
\usepackage{graphics}
\usepackage{rotating}
\usepackage{amsmath}        
\usepackage{amssymb}        
\usepackage{subfigure}        
\usepackage{epsfig}   

\newcommand{\mnras}{MNRAS}
\newcommand{\apjl}{ApJ}
\newcommand{\apj}{ApJ}
\newcommand{\apjs}{ApJS}
\newcommand{\aj}{AJ}
\newcommand{\nat}{Nature}

\newcommand{\aap}{A\&A}

\newcommand{\pasj}{PASJ}

\title[]{The cross-correlation analysis in Z source GX 349+2}

\author[]
{G. Q. Ding$^{1}$\thanks{Email: dinggq@xao.ac.cn, dinggq@gmail.com.}, W. Y. Zhang$^{1,2}$, 
Y. N. Wang$^{3}$, Z. B. Li$^{1}$, J. L. Qu$^{4}$ and C. P. Huang$^{1}$
\\
$^1$Xinjiang Astronomical Observatory, Chinese Academy of Sciences, 150, Science 
1-Street, Urumqi, Xinjiang 830011, China \\
$^2$University of Chinese Academy of Sciences, Beijing 100049, China \\
$^3$Kapteyn Astronomical Institute, University of Groningen, PO BOX 800, NL-9700 AV 
Groningen, the Netherlands \\
$^4$Key Laboratory for Particle Astrophysics, Institute of High Energy Physics, 
Chinese Academy of Sciences, Beijing 100049, China
}

\begin{document}

\date{}



\maketitle

\label{firstpage}

\begin{abstract}

Using all the observations from {\it Rossi X-ray Timing Explorer} for Z source GX 349+2, 
we systematically carry out cross-correlation analysis between its soft and hard X-ray 
light curves. During the observations from January 9 to January 29, 1998, GX 349+2 
traced out the most extensive Z track on its hardness-intensity diagram, making a 
comprehensive study of cross-correlation on the track. The positive correlations and 
positively correlated time lags are  detected throughout the Z track. Outside the Z 
track, anti-correlations and anti-correlated time lags are found, but the anti-correlated 
time lags are much longer than the positively correlated time lags, which might indicate 
different mechanisms for producing the two types of time lags. We argue that neither the 
short-term time lag models nor the truncated accretion disk model can account for the 
long-term time lags in neutron star low mass X-ray binaries (NS-LMXBs). We suggest that 
the extended accretion disk corona model could be an alternative model to explain the 
long-term time lags detected in NS-LMXBs.    

\end{abstract}

\begin{keywords}

binaries: general --- stars: individual (GX 349+2) --- stars: neutron --- 
X-rays: binaries

\end{keywords}

\section{INTRODUCTION}

X-ray binaries (XRBs), consisting of a compact object and a companion star, can be 
divided into low mass X$-$ray binaries (LMXBs) and high mass X$-$ray binaries (HMXBs) 
according to the mass of the companion star. The compact object is either a 
neutron star (NS) or a black hole (BH). Based on their spectral and timing properties, 
NS-LMXBs could be classified as Z sources and atoll sources \citep{Hasinger1989}. 
The Z sources, with high luminosity close to the Eddington limit, trace out a Z-shape 
track on their hardness-intensity diagrams (HIDs), which are divided into three 
branches, called horizontal branch (HB), normal branch (NB), and flaring branch (FB), 
respectively. Among the six confirmed Z sources, Cyg X-2, GX 5-1, and GX 340+0 are 
called Cyg-like Z sources, and the other three Z sources, i.e. Sco X-1, GX 17+2, and 
GX 349+2, are referred to as Sco-like Z sources. On the HIDs of atoll sources with 
luminosity below $\sim$$10^{38}\ {\rm ergs\ s^{-1}}$, two main segments are seen, 
which are called banana state (BS) and island state (IS), respectively; in the BS, 
the variation of hardness is relatively small, while their luminosity spans a wide 
range; however, in the IS, the variation of hardness is obvious, while the change 
of luminosity is relatively small and they are with the lowest luminosity 
\citep{Church2014}. Furthermore, the segment of BS is split into two subsections, 
i.e. the lower banana (LB) with lower luminosities and the upper banana (UB) with 
higher luminosities. In general, atoll sources are either in IS or in BS depending 
on their luminosities. Generally, a Z source cannot become an atoll source and vice 
versa. However, two peculiar NS-LMXBs, i.e. Cir X-1 and XTE J1701-462, show Z source 
behaviors at relatively high luminosities, while they display atoll source behaviors 
at low luminosities \citep{Oosterbroek1995,Shirey1999,Homan2007,Homan2010,Lin2009}.            
                  
The spectral and timing analysis are two main methods for studying XRBs. Among 
the timing analyses, the cross$-$correlation analysis can be used to study the 
relation of light curves in two different energy bands. By analyzing the 
cross$-$correlation function (CCF) of light curves in two different energy bands, 
the correlation and time lag between soft and hard X$-$ray light curves can be 
obtained, which are useful and important to investigate the structures of accretion 
disk and the mechanisms for producing X-rays. According to the behaviors of CCFs, 
the correlations can be classified as anti$-$correlations, positive and ambiguous 
correlations, respectively. Anti$-$correlations correspond to negative 
cross$-$correlation coefficients (CCCs), while positive correlations correspond to 
positive CCCs. If no obvious correlations are presented in the CCFs, such 
correlations are called ambiguous correlations. Soft time lags mean that the lower 
energy photons reach behind, while hard time lags imply that the lower energy 
photons lead the higher energy photons. The cross$-$correlation analysis technique 
has been used to analyze the correlations of X-rays and long-term time lags from 
tens of seconds to over one thousand seconds in two energy bands of XRBs (e.g. 
Choudhury \& Rao 2004; Lei et al. 2008, 2013; Wang, et al. 2014). In addition to 
the CCF method, the cross-spectral analysis is another important method for 
analyzing time lags \citep{Nowak1999a}, which has been widely used to investigate 
the short-term time lags in the order of milliseconds in XRBs (e.g. van der Klis 
et al. 1987; Miyamoto et al. 1988; Cui 1999; Qu, Yu \& Li 2001; Qu et al. 2010a). 
In the past several decades, various time lags have been detected in XRBs through 
analyzing the data from X-ray satellites, of which {\it RXTE} has contributed 
greatly (see review: Poutanen 2001).    

GX 349+2, also known as Sco X-2, is a Sco-like Z source. Using the observations 
from {\it EXOSAT}, \cite{Ponman1988} detected quasi-periodic oscillations (QPOs) 
at $\sim$6 Hz in this source. Using observations from {\it Rossi X-ray Timing 
Explorer (RXTE)}, \cite{ONeill2002} investigated the evolution of QPOs along a 
Z track on its HID and found QPOs at 3.3-5.8 Hz on the FB and at 11-54 Hz 
on the NB. Also through analyzing data from {\it RXTE}, \cite{Zhang1998} detected 
a twin kilohertz (kHz) QPOs with lower and upper frequencies of 712 Hz and 978 Hz, 
respectively. In particular, the kHz QPOs were only found at the top of the NB. 
Using the {\it RXTE} data for this source, \cite{Agrawal2003} carried out the 
spectral evolution on its Z track. They fitted the spectra in 2.5-25 keV using a 
two-component model consisting of a disk blackbody and a Comptonized component 
representing Comptonization in the central hot corona or the boundary layer, 
which could act as the Eastern model \citep{Mitsuda1984,Mitsuda1989}. However, 
\cite{Church2012} used another two-component model consisting of a blackbody and 
a cut-off power law representing Comptonization in an extended corona above the disk 
to fit the spectra of the Sco-like Z sources, including GX 349+2. Using the data 
from {\it BeppoSAX}, \cite{DiSalvo2001} studied its broadband spectra (0.1$-$200 keV) 
and a hard tail was detected in this source. Moreover, they detected an absorption 
edge at $\sim$9 keV in the spectra. Using the observation from XMM, 
\cite{Iaria2009} performed spectral analysis of GX 349+2. They fitted the 
continuum in the 0.7-10 keV energy range with the Eastern model 
\citep{Mitsuda1984}. Significantly, \cite{Iaria2009} found several broad 
emission features below 4 keV and a broader emission feature in the 
Fe-${\rm K_{\alpha}}$ region in the spectra and proposed that these relativistic 
lines are formed due to the reflection in the inner disk region that is 
illuminated by the emission around the NS. The relativistic Fe-K emission line of 
GX 349+2 was also detected in its {\it Suzaku} spectra \citep{Cackett2008}.     

In this work, using all the data from the proportional counter array (PCA) on 
board {\it RXTE} for GX 349+2, we systematically investigate the cross-correlation 
correlations between soft and hard light curves of this source. We describe our 
data analyses in section 2, present our results in section 3, discuss our results 
in section 4, and, finally, give our conclusions in section 5.  

\section{DATA ANALYSIS}

{\it RXTE} made 138 observations from 1996 to 2011 for Z source GX 349+2, 
including 23 observations from January 9 to January 29, 1998, during which the 
source evolved on the most extensive NB+FB tracks ever reported on its HID 
\citep{ONeill2002}. With HEASOFT 6.11 and all the {\it RXTE} observations for 
GX 349+2, we systematically perform cross-correlation analysis for this source. 
In our analysis, only PCA data are needed. The PCA consists of five identical 
proportional counter units (PCUs) in energy range 2$-$60 keV \citep{Jahoda2006}. 
We use the data during the intervals when all five PCUs were working. The PCA 
Standard 2 mode data with bin size of 16 s are used to produce light curves with 
{\it RXTE} FTOOLS SAEXTRCT. Then, with RUNPCABACKEST, a {\it RXTE} script, we 
produce the background files from the bright background model 
(pca$_{-}$bkgd$_{-}$cmbrightvle$_{-}$eMv20051128.mdl) provided by {\it RXTE} 
team and thus produce the background light curves. Finally, applying LCMATH, a 
XRONOS tool, we generate the background-subtracted light curves with various 
energy bands. When extracting light curves, good time intervals (GTIs) are 
restricted through inputting GTI files, which are created with the FTOOLS 
MAKETIME obeying criteria: the earth elevation angle greater than $10^{\circ}$ 
and the pointing offset less than $0.02^{\circ}$.      

\begin{figure}
\vspace{5.cm}
\centerline{
\includegraphics[width=13.cm, height=10.cm]{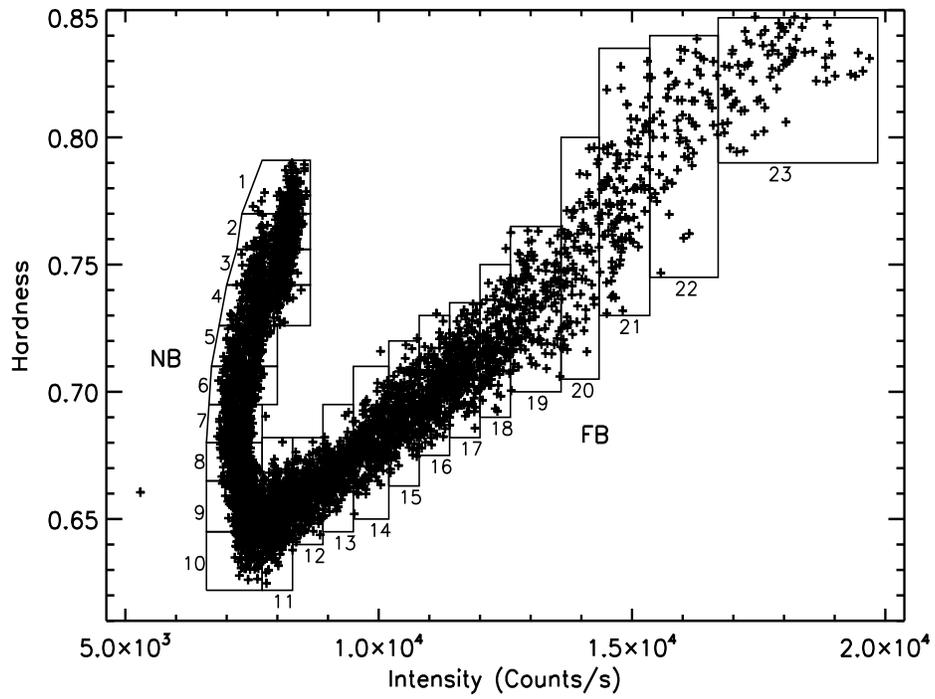}}
\vspace{0.25cm}
\caption{The HID of GX 349+2. Each point represents 16 s background-subtracted data. The 
hardness is defined as the count rate ratio between 8.7-19.7 keV and 6.2-8.7 keV energy 
bands, and the intensity as the count rate in the 2.0-19.7 keV energy band. In orde to 
investigate cross-correlation evolution on the HID, the track of the extensive Z track is 
divided into 23 regions (labeled 1, 2, ..., 23) that are used to group data for production 
of CCF of each region.}\label{fig:HID}
\end{figure}

To build the HID on which the source traced out the most extensive Z track, 
following \cite{ONeill2002}, we define the hardness as the count rate ratio 
between 8.7$-$19.7 keV and 6.2$-$8.7 keV energy bands, and the intensity as the 
count rate in the 2.0$-$19.7 keV energy band. The produced HID is shown by 
Figure~\ref{fig:HID}. In order to investigate the evolution of cross-correlation 
correlations along the Z track, we divide the track into 23 regions. For minimizing 
the variation of count rate and meanwhile having enough observation time in each 
region, these regions are produced obeying the criteria: the count rate variation 
in each region is less than 1500 count $s^{-1}$ except the region No. 23; the 
hardness variation in each of the regions in the NB and in the first half FB is 
less than 0.025, while it is less than 0.5 for each region in the second half FB, 
because, here, the points are scattered. Then, we study the evolution of 
cross-correlation correlations between hard and soft X-ray light curves 
along the Z track. Firstly, based on the values of hardness and count rate in 
each region, we determine the relative and absolute time intervals of each region 
with FTOOLS MAKETIME and TIMETRANS, respectively. Secondly, inputting the absolute 
time intervals of each region when light curves are extracted, we produce the 
soft and hard X-ray background-subtracted light curves of each region in 2-5 keV 
and 16-30 kev energy bands, respectively. Thirdly, with XRONOS TOOL CROSSCOR, 
we generate the CCF between the soft and hard background-subtracted light curves 
of each region. To get the CCCs and time lags, we fit the CCFs with an inverted 
Gaussian function at a 90\% confidence level. The results of the 23 regions are 
listed in Table~\ref{table1}. The total relative time length of each of the 23 
regions spans from 1072 s to 8606 s and the absolute time intervals of each region 
spread within 2 days. 

\begin{figure}
\vspace{1.cm}
\centerline{\hbox{
\includegraphics[width=8cm, height=5cm]{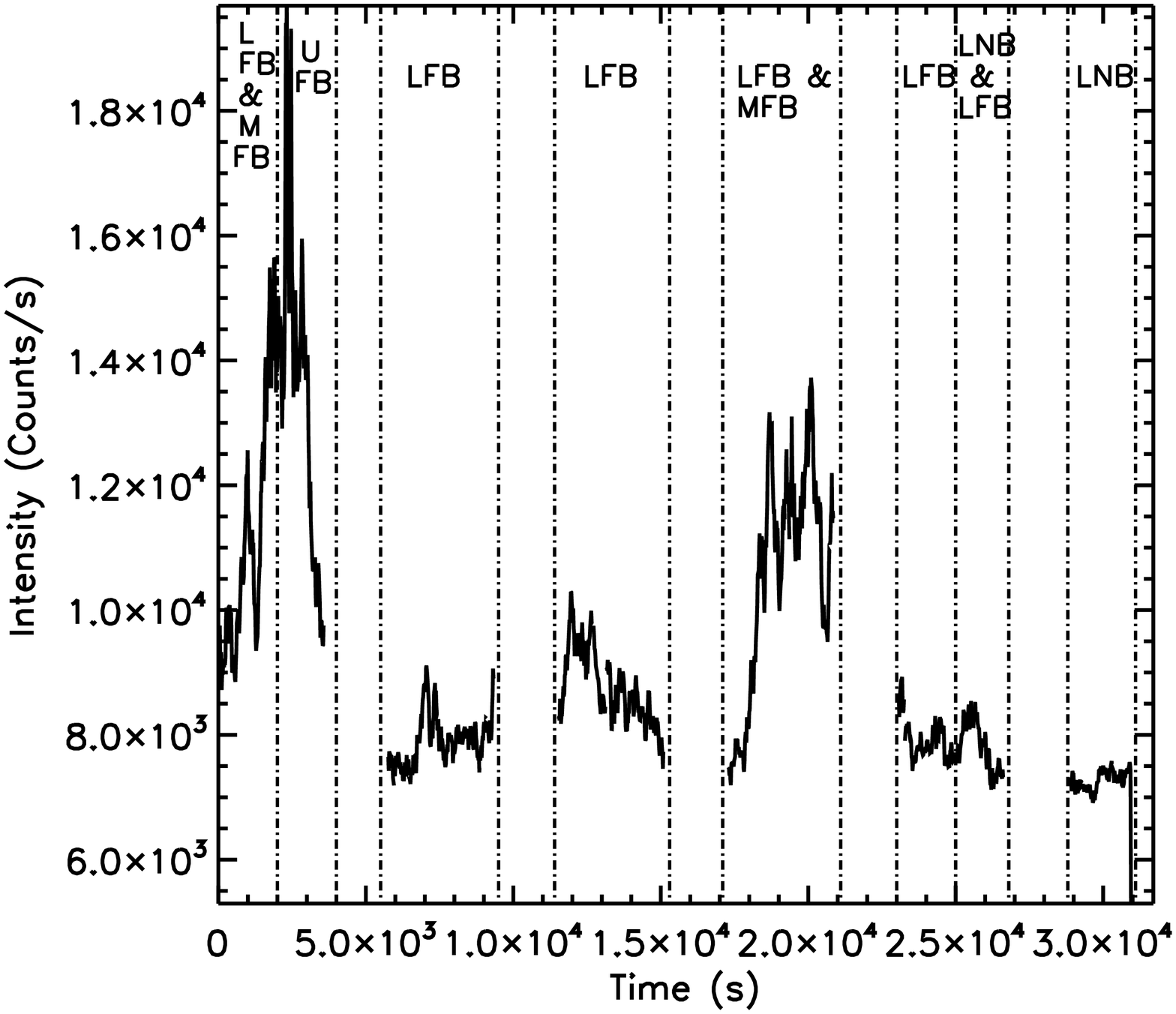}
\hspace{-0.15cm}
\includegraphics[width=8cm, height=5cm]{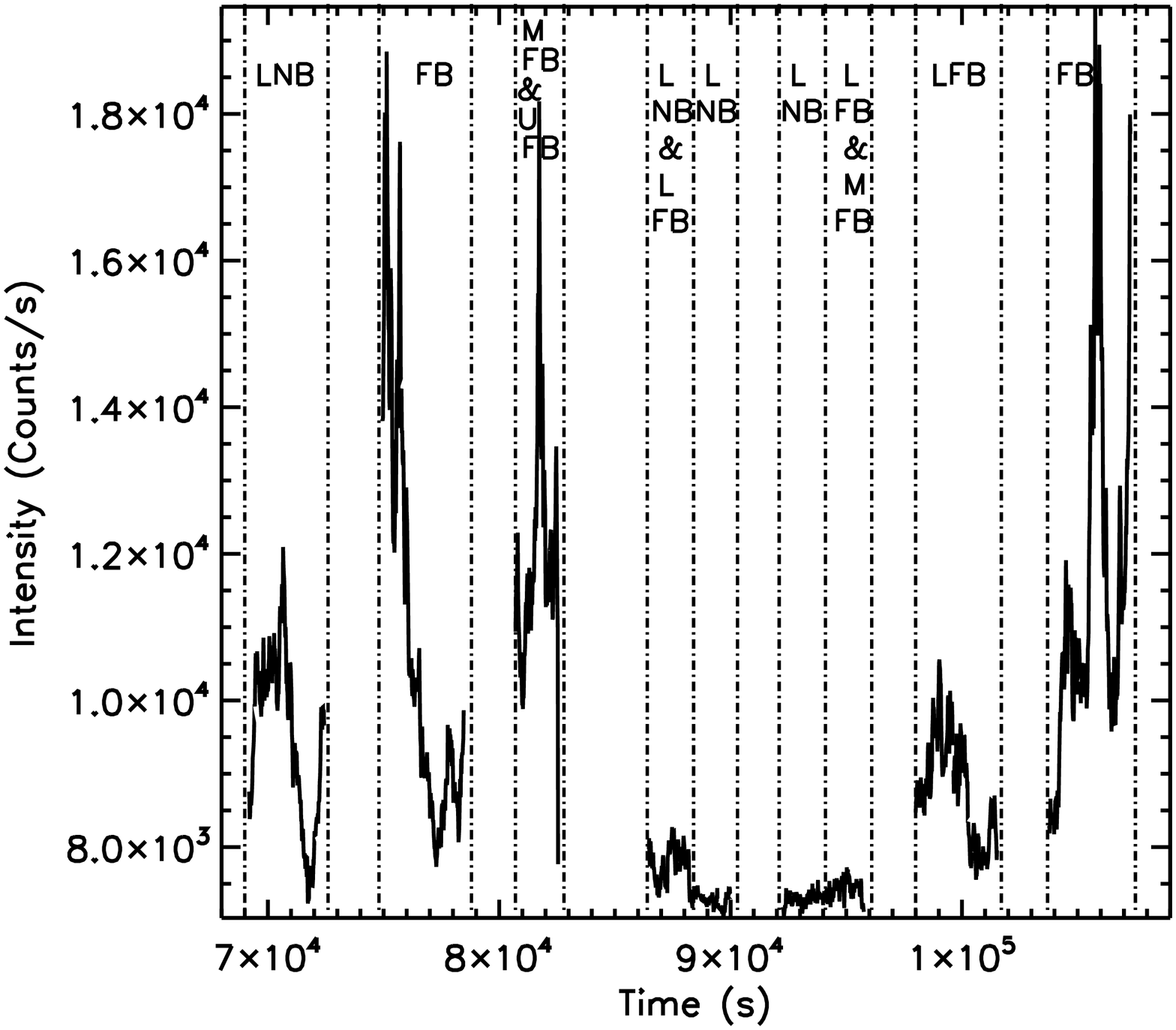}
}} 
\vspace{-0.1mm}
\centerline{\hbox{
\includegraphics[width=8cm, height=5cm]{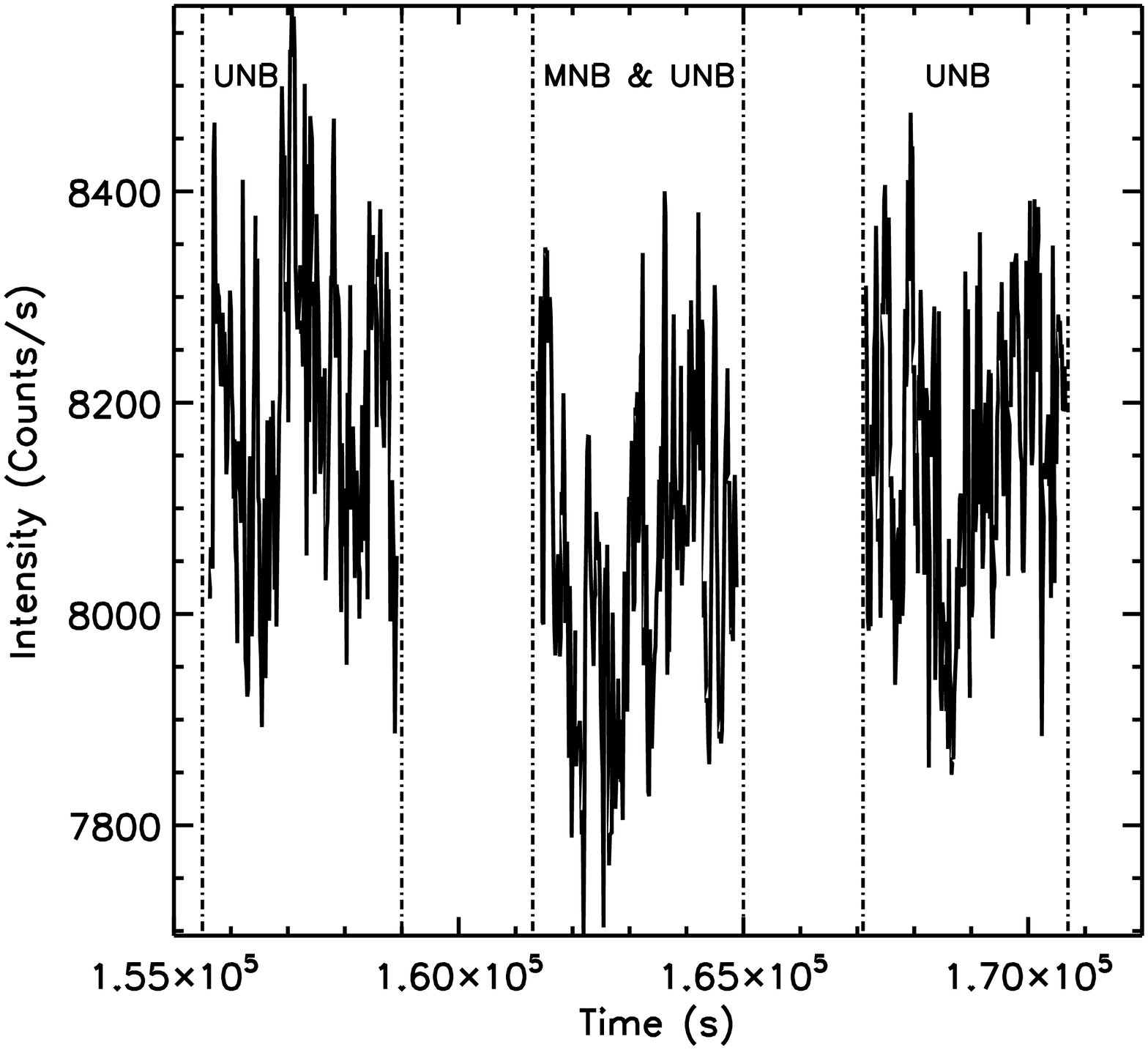}
\hspace{-0.15cm}
\includegraphics[width=8cm, height=5cm]{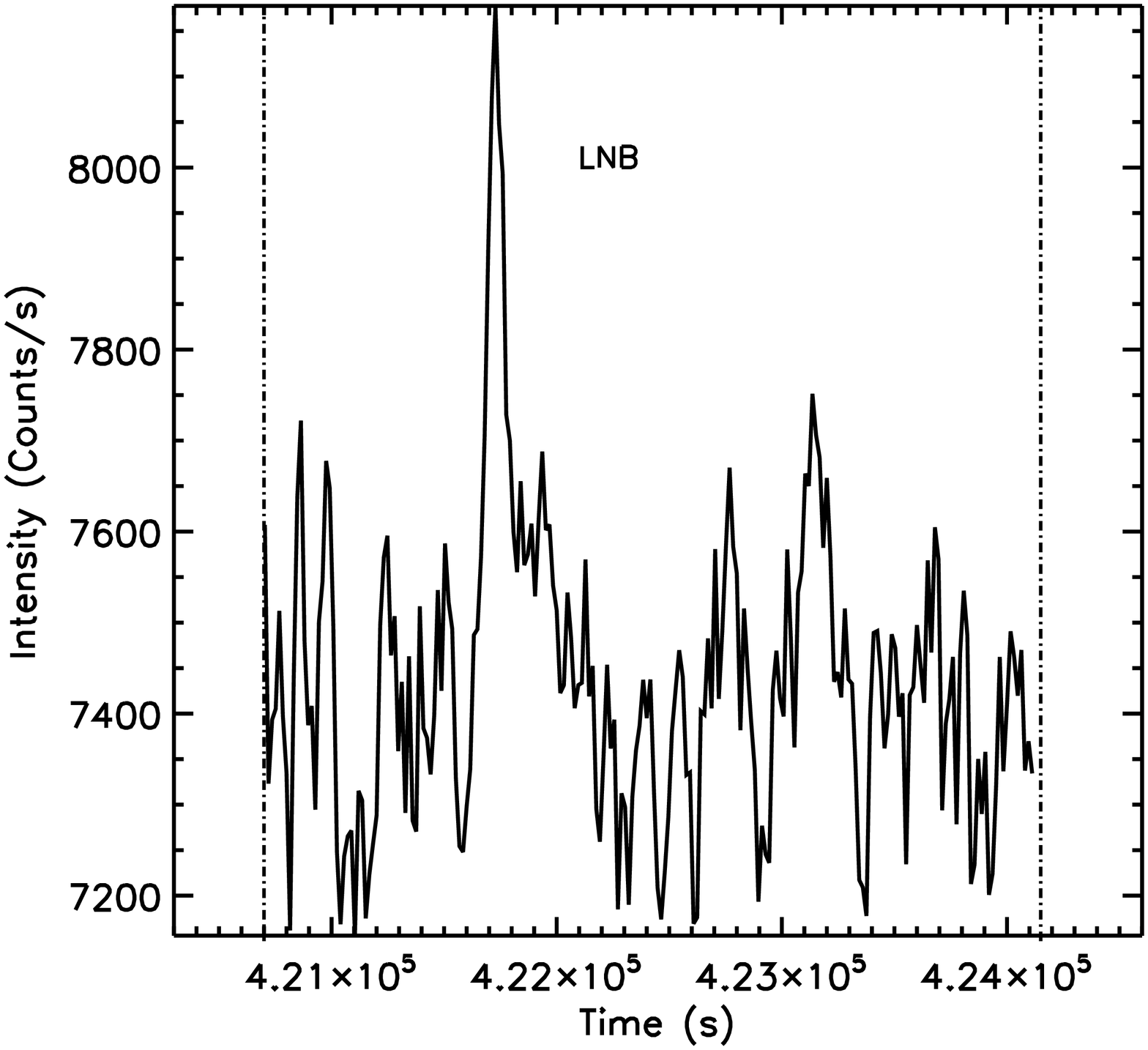}
}} 
\vspace{-0.1mm}
\centerline{\hbox{
\includegraphics[width=8cm, height=5cm]{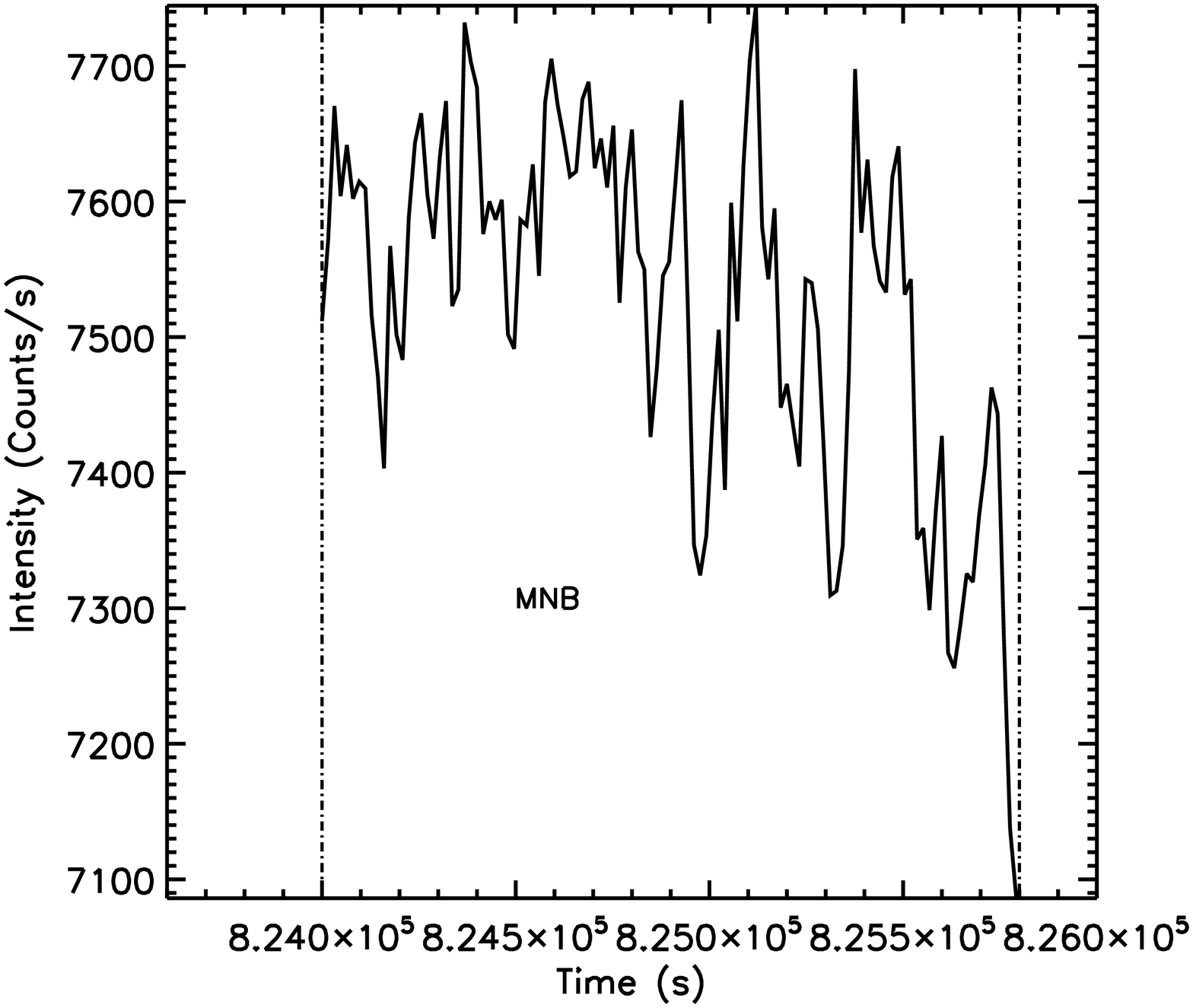}
\hspace{-0.15cm}
\includegraphics[width=8cm, height=5cm]{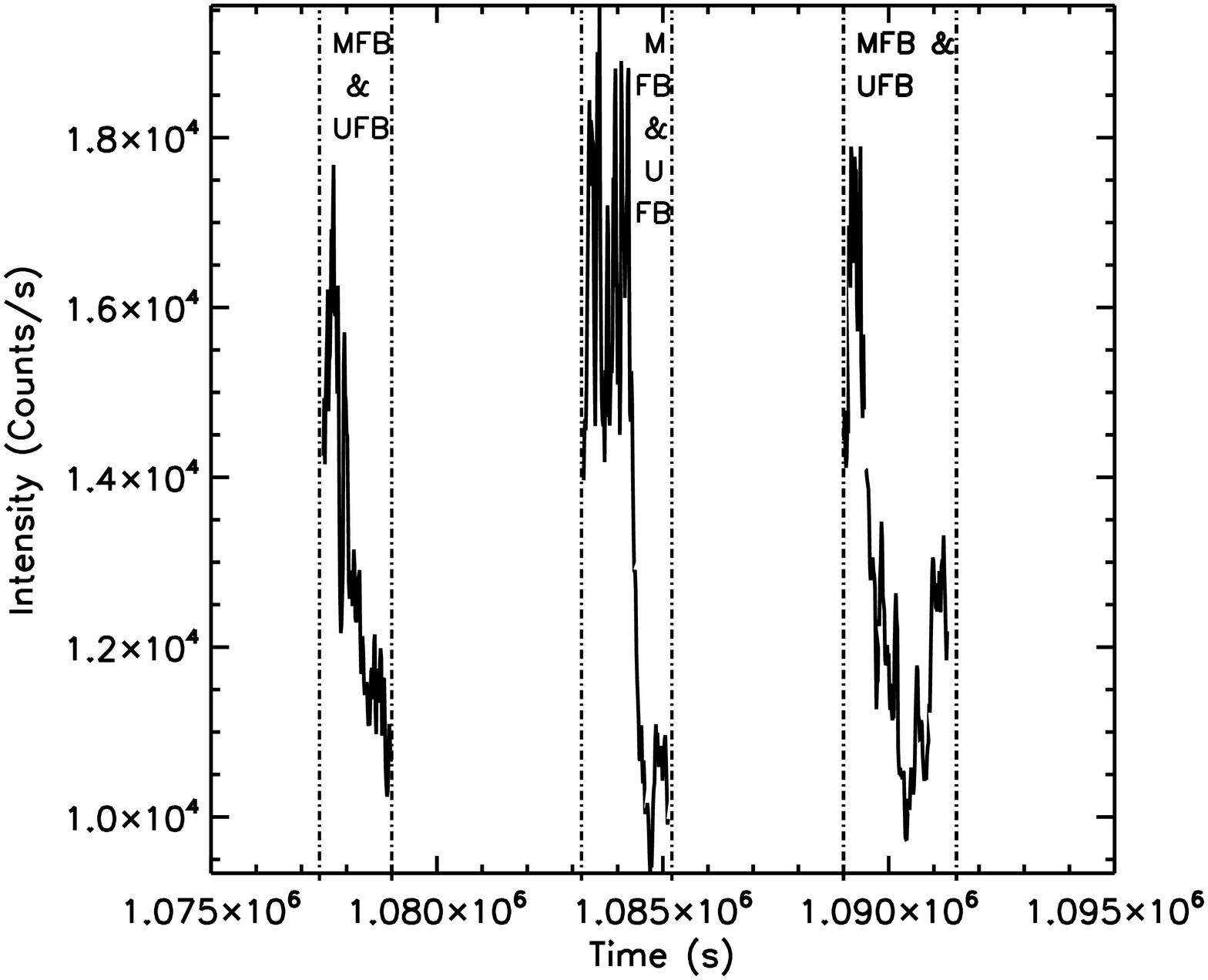}
}} 
\vspace{-0.1mm}
\centerline{\hbox{
\includegraphics[width=8cm, height=5cm]{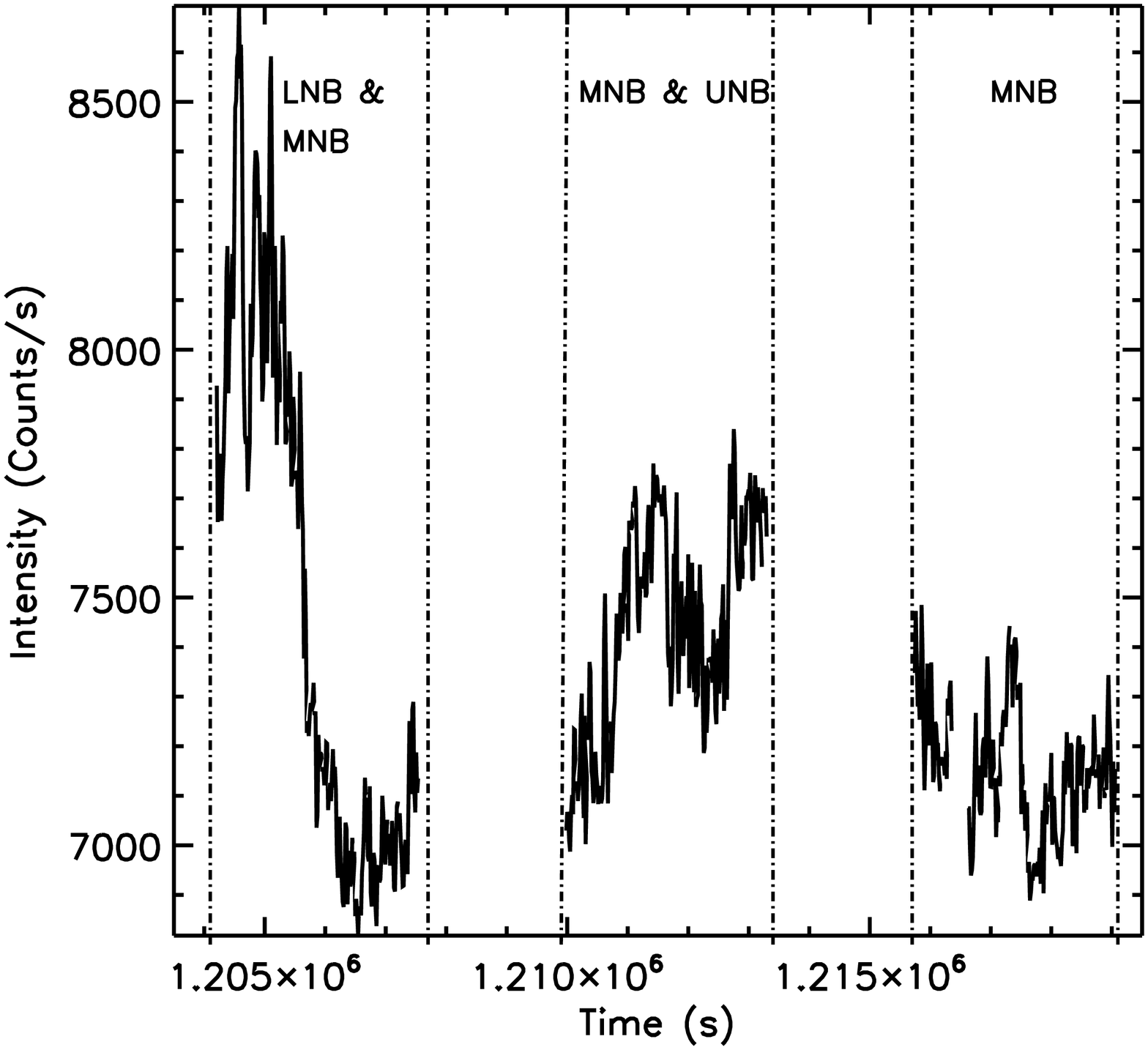}
\hspace{-0.15cm}
\includegraphics[width=8cm, height=5cm]{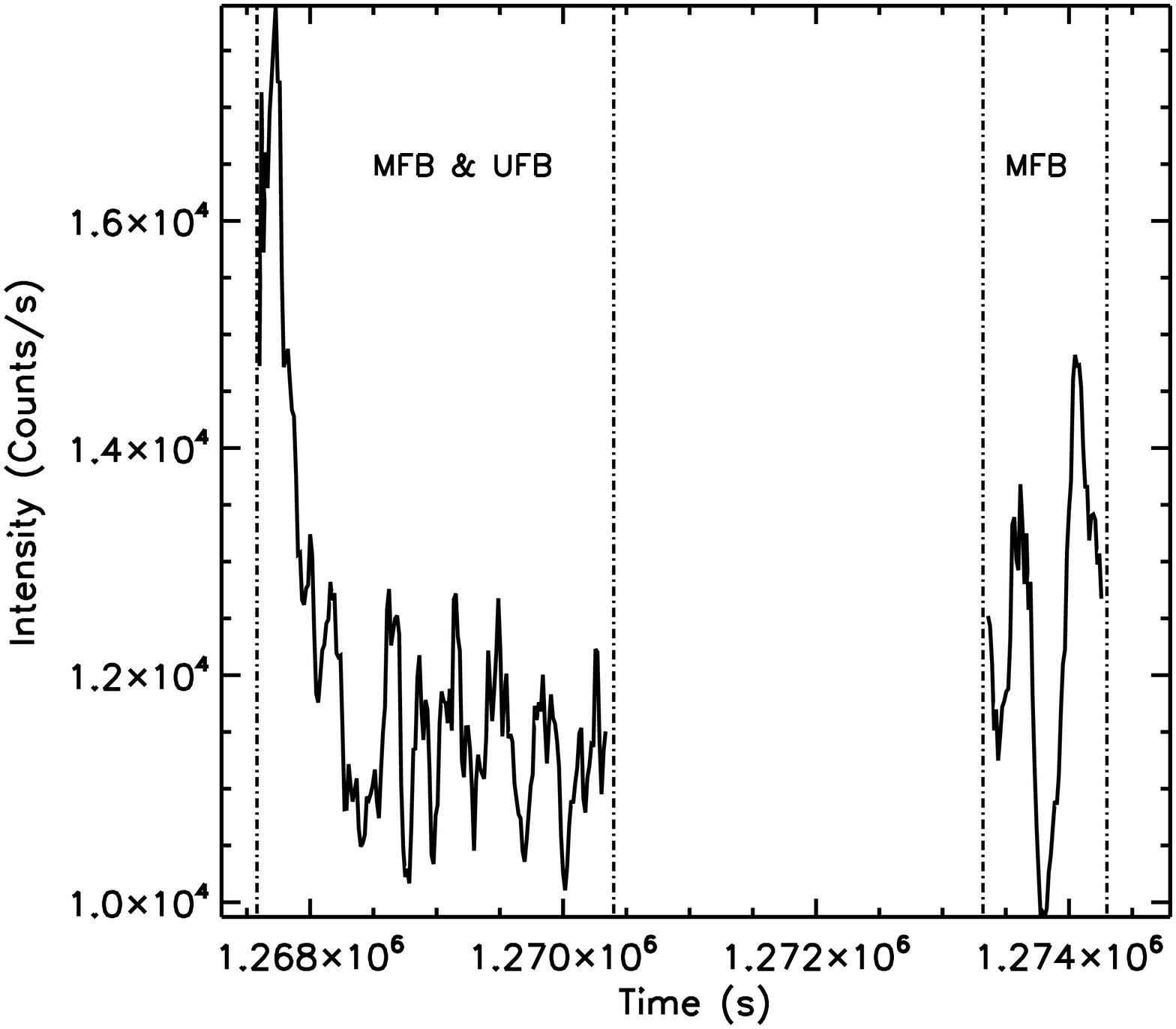}
}} 
\caption{The light curve during the period tracing out the HID. The positions of segments 
on the HID are identified. The light curve shows that on the HID the source evolves from the 
FB to the NB and it shows two evolutionary cycles and the beginning of the third 
evolutionary cycle.}\label{fig:lightcurve}
\end{figure}

As shown in Figure~\ref{fig:HID}, the track consists of an extensive NB and an 
elongated FB, respectively. Regions 1-10 constitute the NB, while regions 11-23 
make up the FB. In order to describe the positions of the 23 regions on the 
NB+FB tracks, the segment of regions 1-3 is called upper NB (UNB), the segment 
of regions 4-6 is called middle NB (MNB), and the segment of regions 7-10 is 
called lower FB (LFB); similarly, the segment of regions 11-15 is named lower 
FB (LFB), the segment of regions 16-18 is named middle FB (MFB), and the segment 
of regions 19-23 is named upper FB (UFB). The intensity as a function of time 
during the HID is shown in Figure~\ref{fig:lightcurve}, in which the positions 
of various time intervals on the HID are marked. As shown in 
Figure~\ref{fig:lightcurve}, the two panels of the first row dominate the FB 
positions, while the two panels of the second row and the left panel of the 
third row are occupied by the NB positions, showing that the source
evolves from the FB to the NB on the NB+FB tracks; the right panel of the third 
row and the left panel of the fourth row show the next evolutionary cycle from 
the FB to the NB; the right panel of the fourth row shows the beginning of the 
third evolutionary cycle. Therefore, the source regularly evolved on the HID 
during each evolutionary cycle, which leads to that the data of each region 
spread within two days. For Fourier analysis, unbroken sections of data are 
needed. Certainly, for cross-correlation analysis continuous observations should 
be better than broken observations. Here, although the observations within the 
HID span 21 days, yet the data within each region do not spread across very 
different observations, which ensures the validity of our results.

In addition, we perform cross-correlation analysis between soft and hard X-ray 
light curves for all the observations outside the period during which the 
source traced out the most extensive NB+FB tracks on its HID. Similarly, we 
produce the soft and hard X-ray background-subtracted light curves of each 
observation and then get its CCF, CCC, and time lag. It is noted that if 
there are several segments in the light curves of an observation, we 
produce the CCF of each segment.        
            
\section{RESULTS}

\begin{figure}
\vspace{5.cm}
\centerline{\hbox{
\includegraphics[width=8cm, height=6cm]{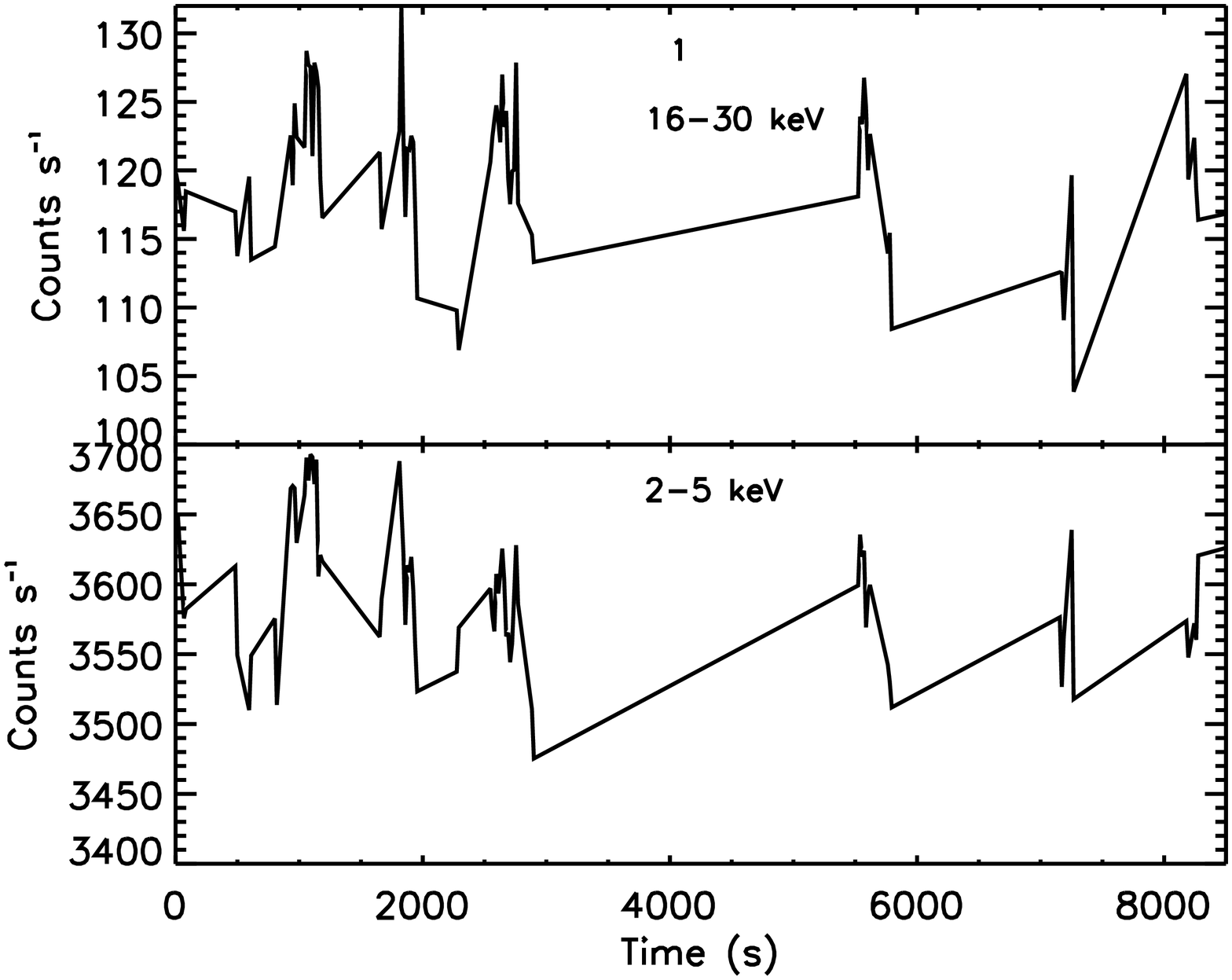}
\hspace{-0.15cm}
\includegraphics[width=8cm, height=6cm]{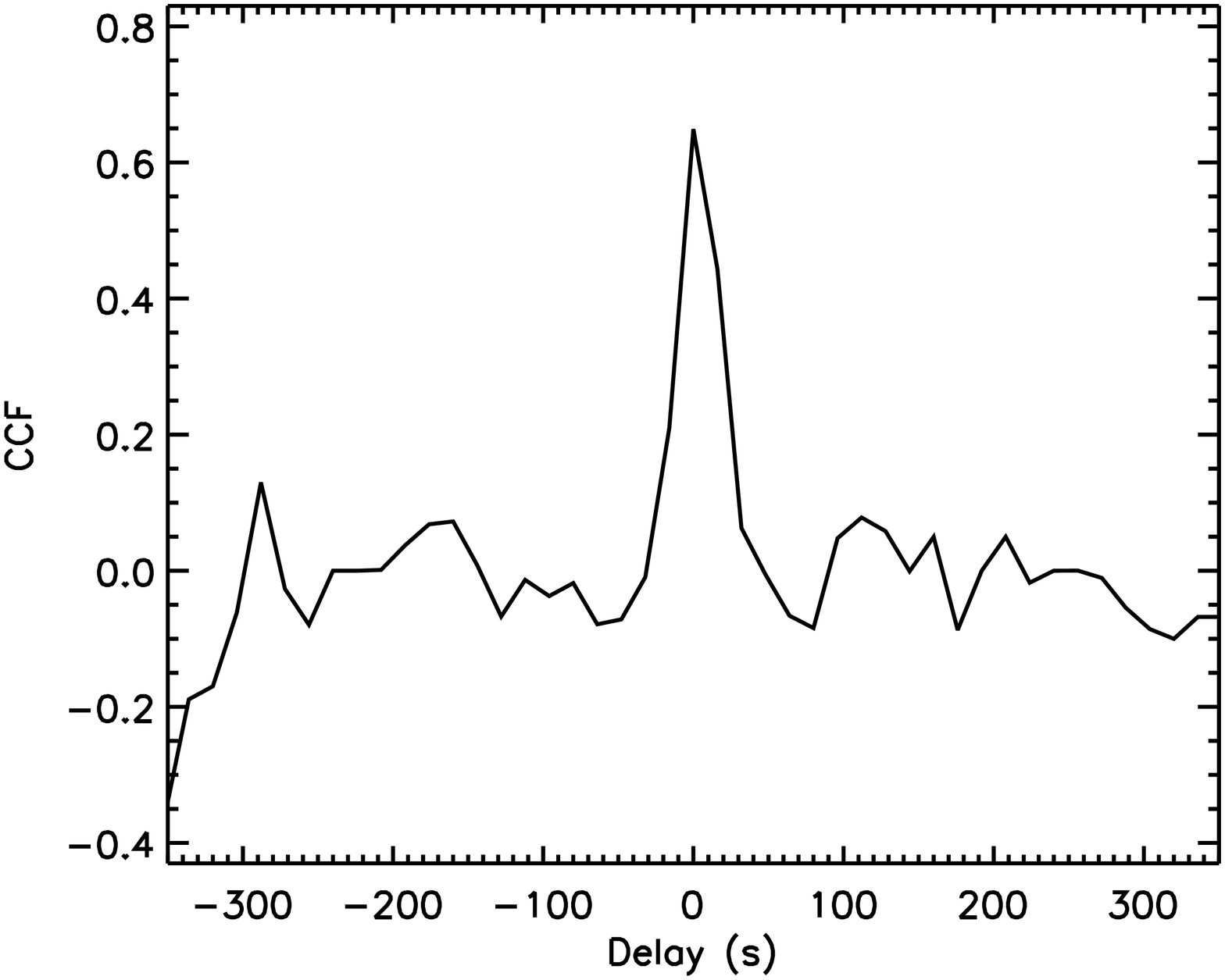}
}} 
\vspace{-0.1mm}
\centerline{\hbox{
\includegraphics[width=8cm, height=6cm]{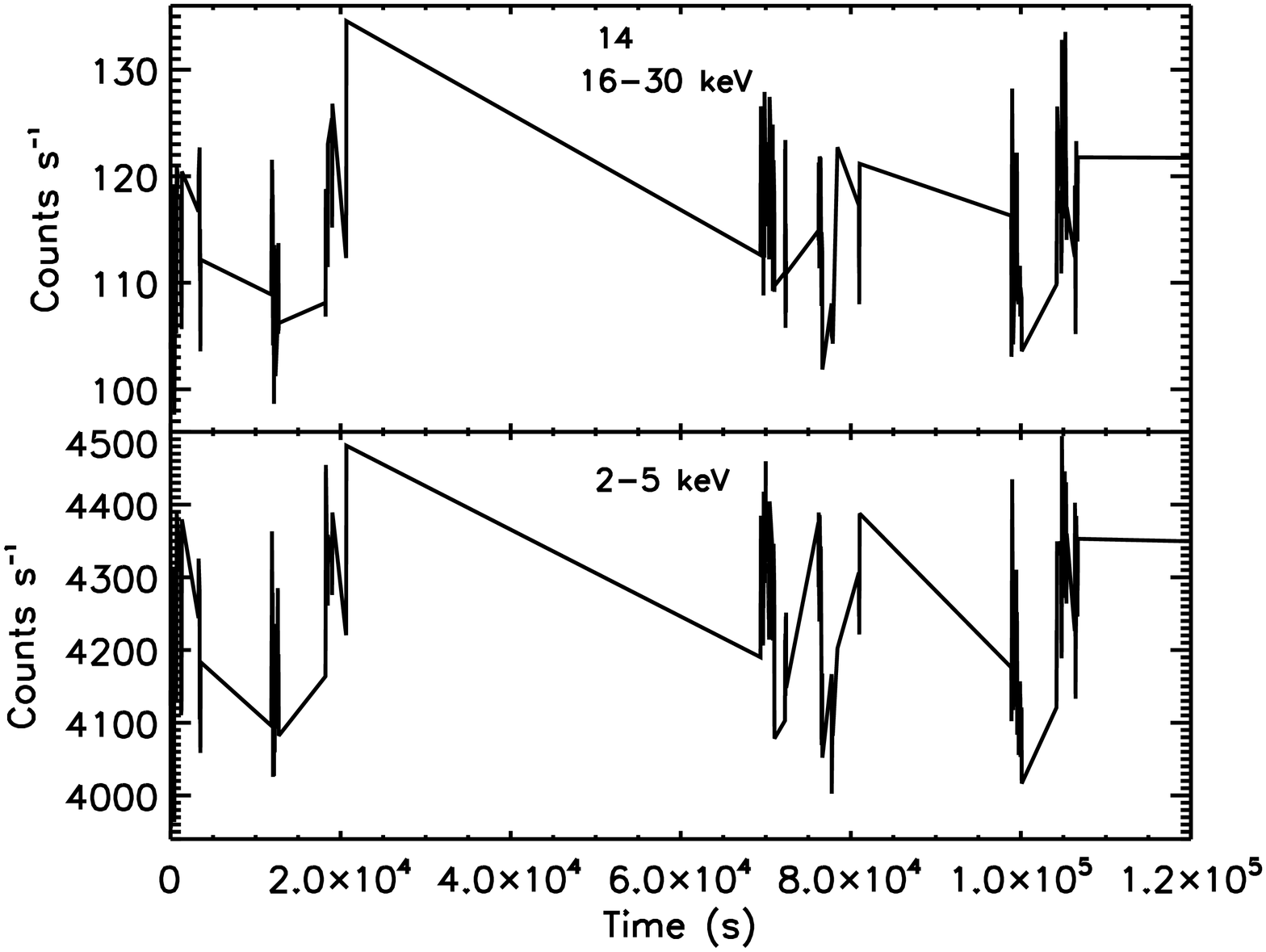}
\hspace{-0.15cm}
\includegraphics[width=8cm, height=6cm]{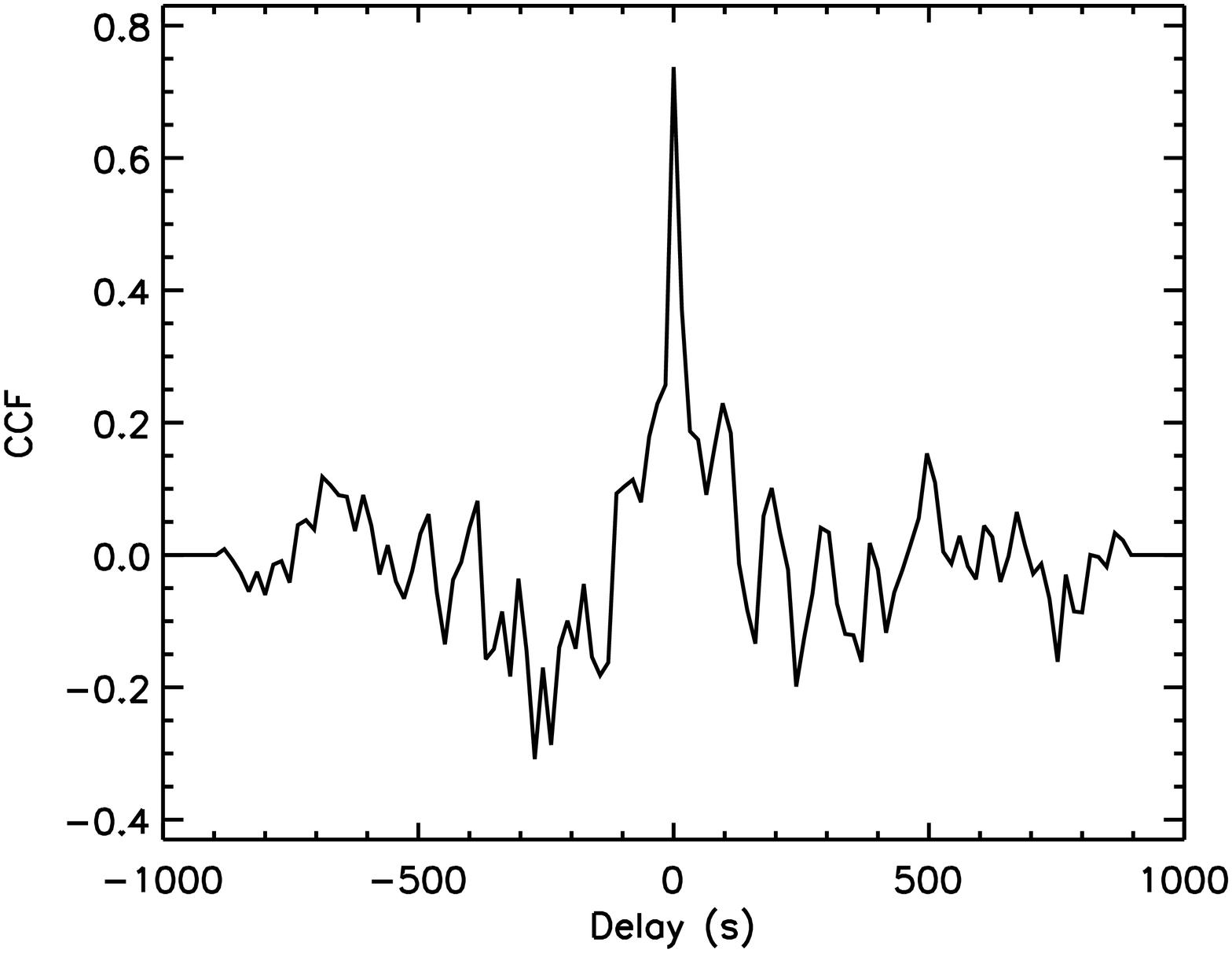}
}} 
\caption{Left panels: the hard light curves (16-30 keV) and soft light 
curves (2-5 kev) of two representative HID regions (1, 14) in which 
positive correlations are detected. Right panels: the CCFs of the two regions.  
}\label{fig:positive}
\end{figure}

\begin{figure}
\vspace{5.cm}
\centerline{\hbox{
\includegraphics[width=8cm, height=6cm]{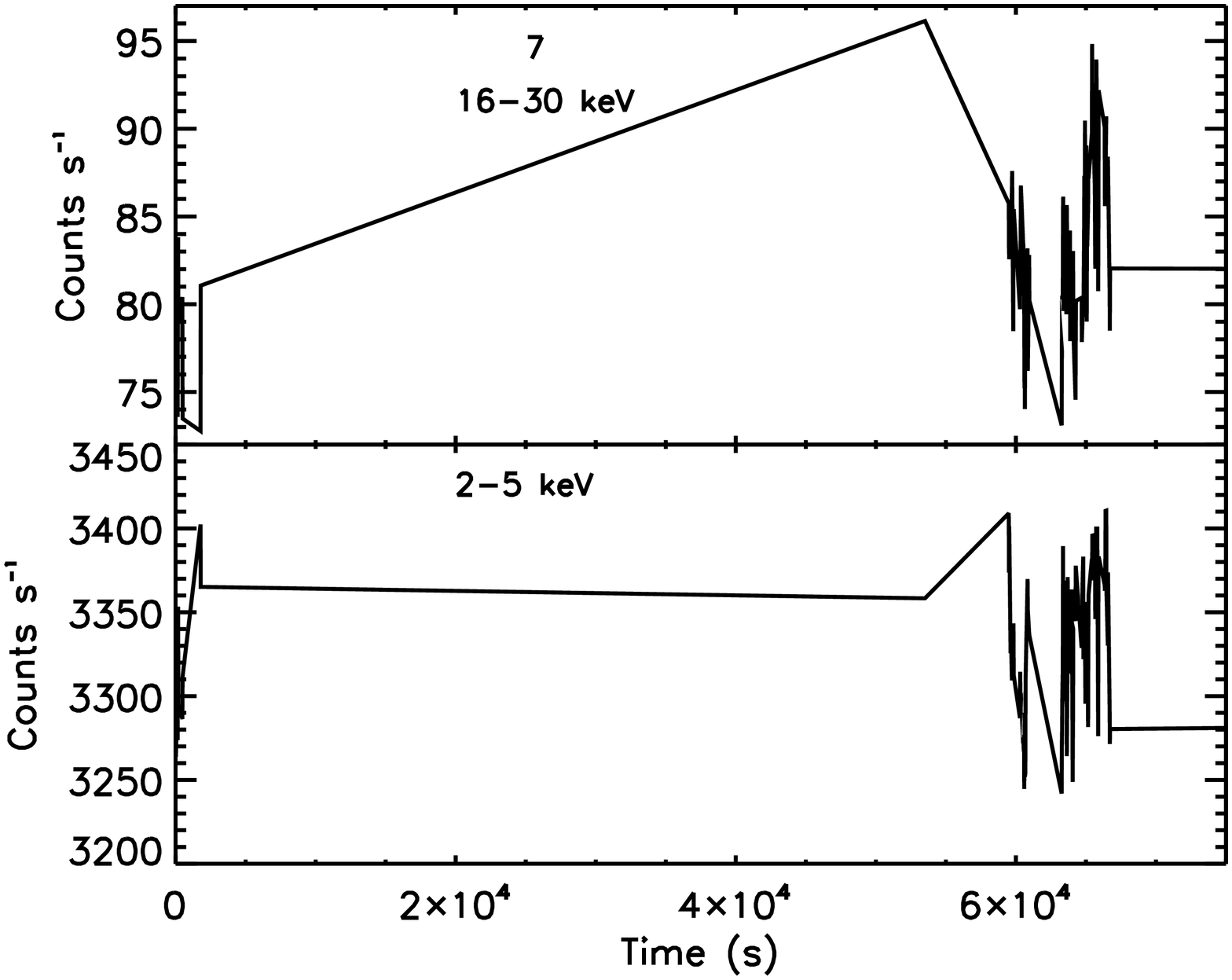}
\hspace{-0.15cm}
\includegraphics[width=8cm, height=6cm]{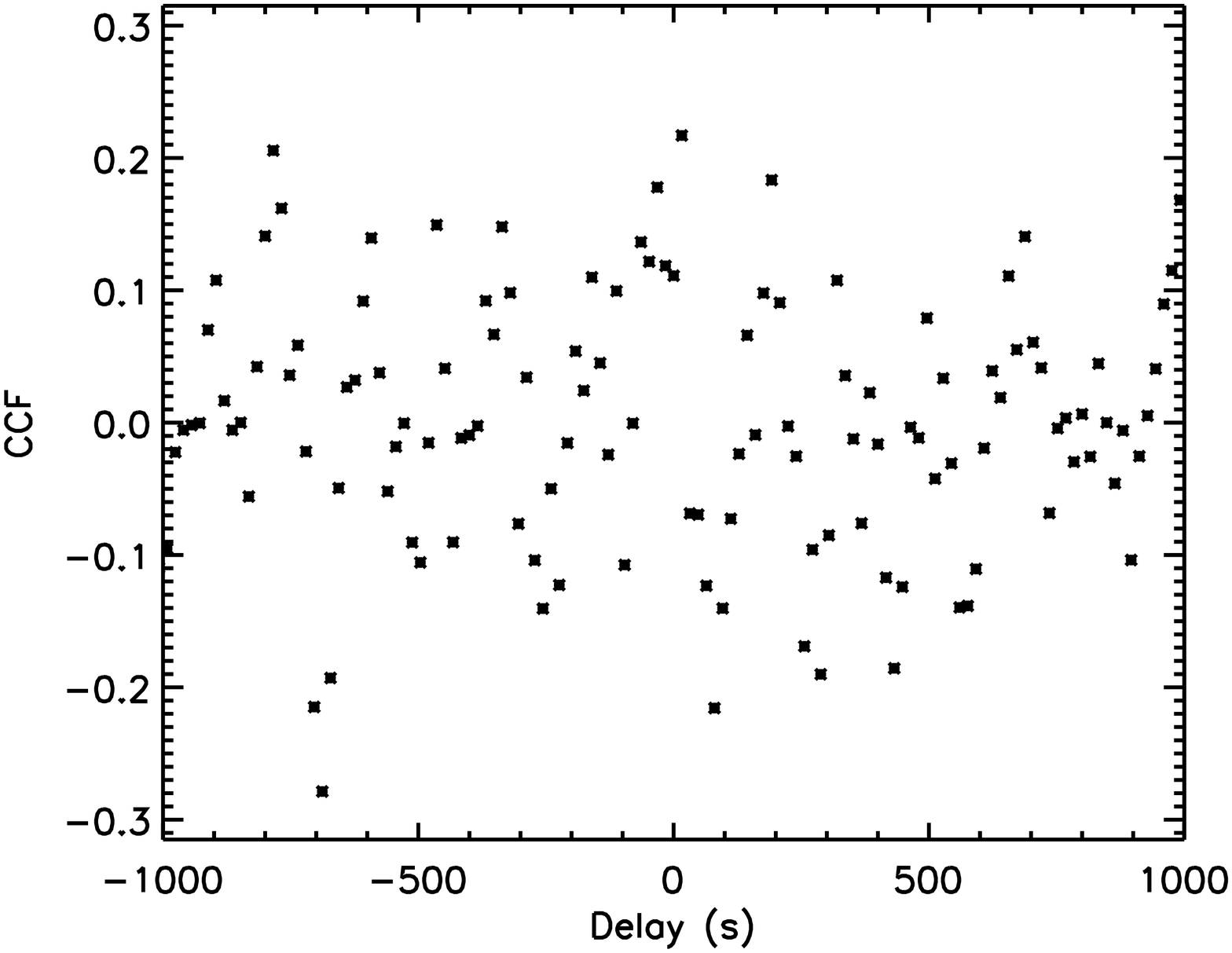}
}} 
\caption{The upper and low panels on the left show the hard X-ray light curve (16-30 keV) 
and soft X-ray light curve (2-5 keV) of region 7 of the HID, respectively. The panel on the 
right shows the CCF of the two light curves, where a typical ambiguous correlation is shown.
}\label{fig:ambiguous}
\end{figure}

\begin{figure}
\vspace{5.cm}
\centerline{\hbox{
\includegraphics[width=8cm, height=6cm]{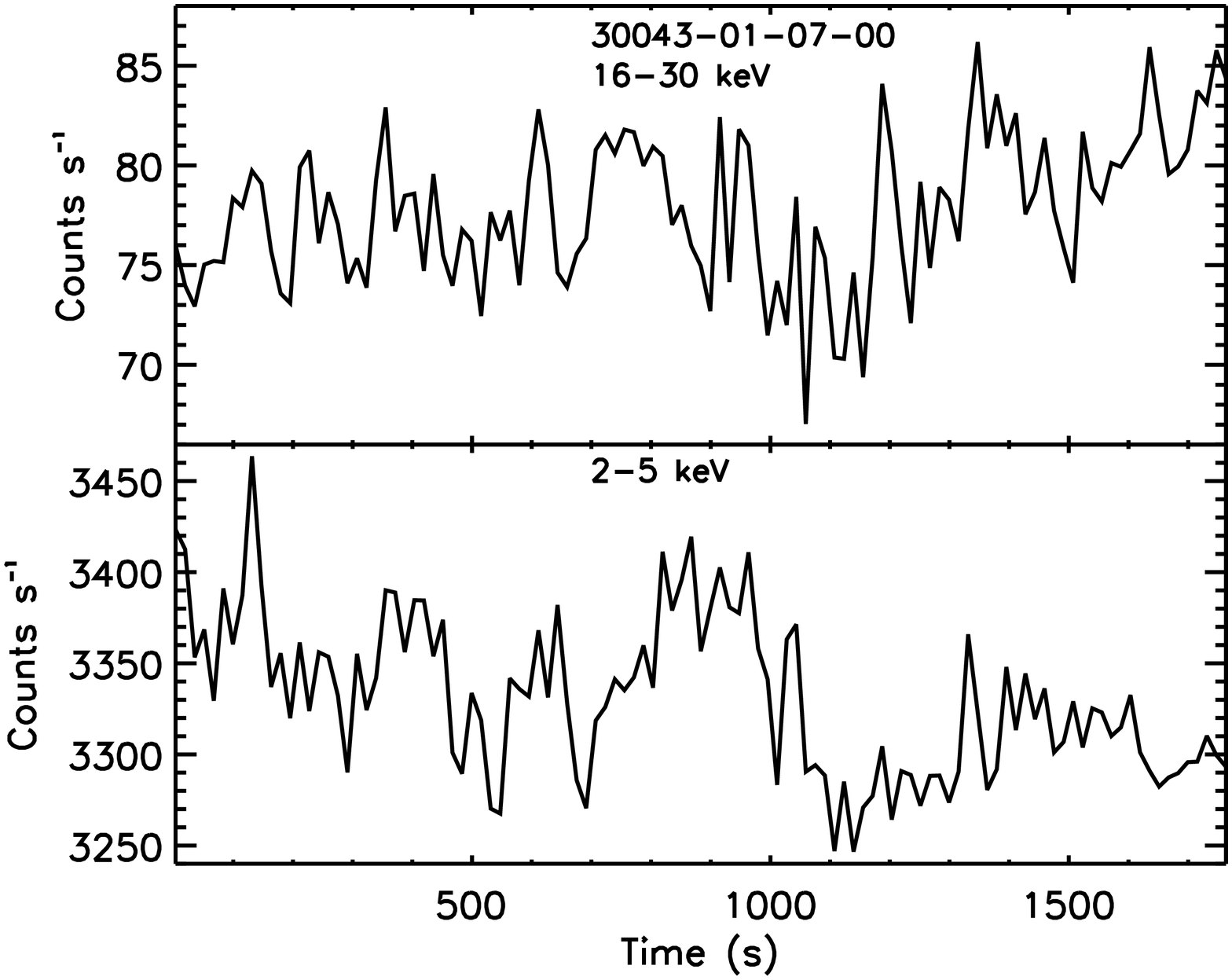}
\hspace{-0.15cm}
\includegraphics[width=8cm, height=6cm]{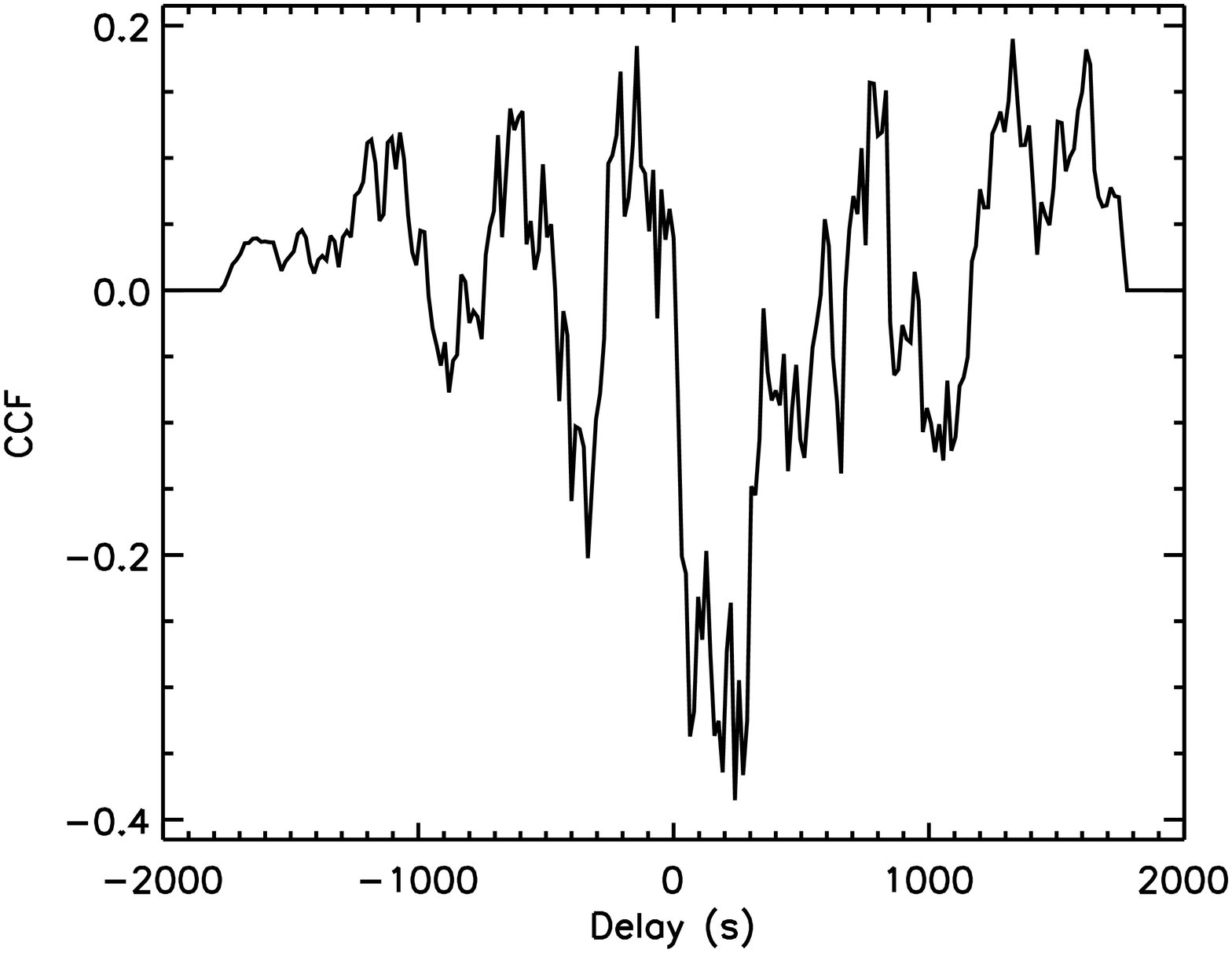}
}} 
\vspace{-0.1mm}
\centerline{\hbox{
\includegraphics[width=8cm, height=6cm]{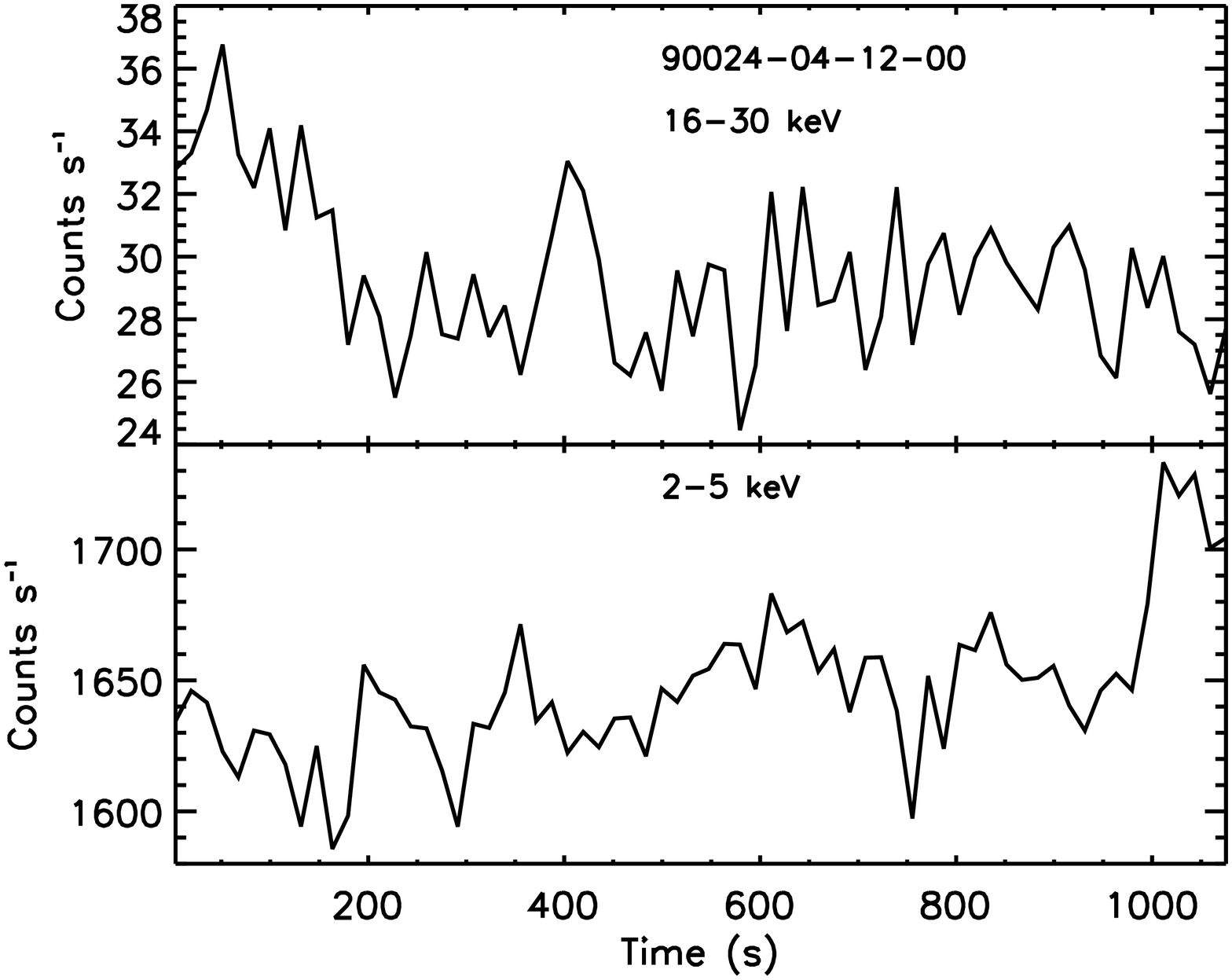}
\hspace{-0.15cm}
\includegraphics[width=8cm, height=6cm]{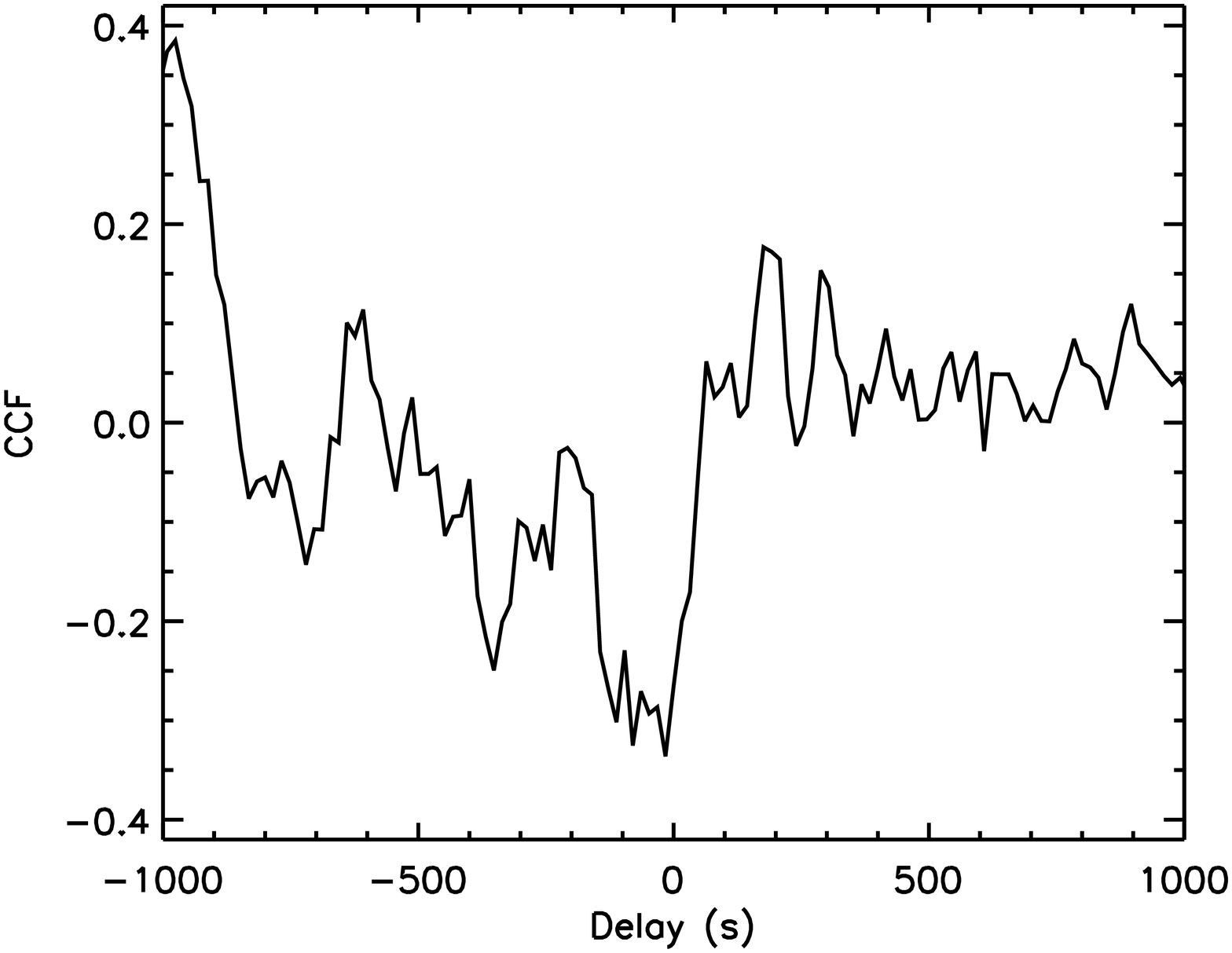}
}} 
\caption{Two representative anti-correlated correlations which are detected in the epoch 
outside the period tracing out the HID. Left panels: the hard X-ray light curves (16-30 keV) and 
soft X-ray light curves (2-5 kev) of two observations in which anti-correlated correlations are 
detected; right panels: the CCFs of the two observations.}\label{fig:anti-correlation}
\end{figure}

In this work, using all the 138 {\it RXTE} observations from 1996 to 2011 
for Z source GX 349+2, we systematically perform cross-correlation analysis 
between 2-5 kev and 16-30 keV light curves with the CCF method. The source 
traced out the most extensive Z track on the HID during the 23 observations 
from January 1 to January 29, 1998. Among the 23 regions on the HID, positive 
correlations and ambiguous correlations are detected in 18 and 5 regions, 
respectively, while anti-correlations are not found throughout the Z track. 
The HID positions of the 18 regions, the derived CCCs and time lags, and 
hardness values are listed in Table~\ref{table1}. Figure~\ref{fig:positive} 
and Figure~\ref{fig:ambiguous} show two detected positive correlations and 
one ambiguous correlation, respectively. Eight regions of the 18 regions with 
detected positive correlations are assigned to the NB and other ten regions 
are belonged to the FB. Among the five regions with ambiguous correlations, 
the NB and FB host two and three regions, respectively. Fortunately, among 
the observations outside the period during which the source traced out the 
most extensive Z track on its HID, the anti-correlations are detected in ten 
observations. The analysis results of these anti-correlations are listed in 
Table~\ref{table2} and two representative anti-correlations are shown in 
Figure~\ref{fig:anti-correlation}. The derived anti-correlated time lags vary 
between tens of seconds and thousands of seconds. Among the ten observations 
with anti-correlations, hard and soft X-ray time lags are detected in four and 
six observations, respectively. In one hundred observations outside the HID 
period, positive correlations are detected, which are listed in Table~\ref{table3}, 
while ambiguous correlations are found in five observations outside the HID 
episode. The anti-correlations, positive correlations, and ambiguous correlations 
are detected in 8.7\%, 87\%, and 4.3\% of the total observations outside the HID 
period, respectively. Comparing the time lags listed in 
Tables~\ref{table1}-\ref{table3}, one can see that the positively correlated 
time lags vary from several seconds to tens of seconds, whereas the 
anti-correlated time lags span a wide range from tens of seconds to thousands of 
seconds, so the anti-correlated time lags of GX 349+2 are much larger than its 
positively correlated time lags, which is consistent with what was found in 
atoll source 4U 1735-44 and 4U 1608-52, as well as peculiar source XTE J1701-462 
\citep{Lei2013,Lei2014,Wang2014}. It should be informed that a small number of 
short-term time lags less than one second are obtained in the positive 
correlations of GX 349+2, as listed in Tables~\ref{table1} and \ref{table3}.

\section{DISCUSSION}

\subsection{The short-term time lags in XRBs}

Through cross-spectral analysis, the short-term time lags ($<$1 s) between the 
emissions of two adjacent X-ray energy bands have been detected in a few Galactic 
black hole X-ray binaries (BHXBs) and NS-LMXBs. Among the BHXBs in which short-term 
time lags were studied, the short-term time lag behaviors of Cyg X-1 were investigated 
most deeply \citep{Miyamoto1988,Miyamoto1992,Crary1998,Nowak1999a,Pottschmidt2000,Negoro2001}. 
The short-term time lags were also detected in microquasar GRS 1915+105 
\citep{Cui1999,Reig2000,Pahari2013}, in GX 339-4 \citep{Miyamoto1992,Nowak1999b}, as well 
as in black hole candidate GS 2013+338 and GRO J0422+32 \citep{Miyamoto1992,vanderHooft1999}. 
The short-term time lags with values of several milliseconds were also found in NS-LMXBs, 
such as in Cyg X-2 and GX 5-1 \citep{vanderKlis1987,Vaughan1994} as well as in Cir X-1 
\citep{Qu2001,Qu2010b}. Generally, the hard X-ray short-term time lags were detected in the 
hard states of BHXBs, while both hard and soft X-ray short-term time lags were detected in 
NS-LMXBs \citep{vanderKlis1987,Qu2001}. 

Some models have been proposed to explain short-term time lags, but the observed various 
short-term time lags cannot be explained by any single model. The cross-spectral analysis 
technique is based on the Fourier transform, so the time lag spectrum, i.e. the correlation 
between the short-term time lag and the frequency of Fourier transform can be obtained. 
Usually, the short-term time lag is anti-correlated with Fourier frequency \citep{vanderKlis1987,vanderHooft1999,Nowak1999a,Nowak1999b,Pottschmidt2000,Qu2010a}. Among the 
Fourier frequencies, some are the frequencies of QPOs, sequentially leading to that the 
short-term time lag is also anti-correlated with QPO frequency. Since the short-term time 
lag is in connection with the QPO parameter, it could be feasible to invoke the models 
responsible for QPOs to account for the short-term time lags, such as the shot model 
\citep{Terrell1972}. In the shot model, it is assumed that the gravitation energy of 
accreting matter is transformed to thermal energy emitted as successive bursts, which 
are called ``shots'' and responsible for QPOs; the different shot profiles or shot 
distribution in different energy bands results in the time delays between the hard and 
soft X-ray emissions \citep{Miyamoto1989,Nowak19994,Nowak1999b,Poutanen2001}. Although 
the shot model provided an explanation for the strong aperiodic variability of the flux 
of accreting XRBs immediately after such variability was discovered and it was very 
early applied to account for the short-term lags, yet this model inevitably encountered 
difficulty when it was used to explain the large X-ray variability timescale 
range \citep{Uttley2001,Churazov2001,Revnivtsev2009} and the linear rms-flux relation 
\citep{Uttley2001,Ingram2013}. Since the shot model, various models have been proposed to 
explain the QPOs in XRBs, such as the beat-frequency 
models \citep{Strohmayer1996,Ford1997,Miller1998}, which were refuted by the fact that the 
observed QPO peak separation is not a constant \citep{vanderKlis2000}, the Lense-Thirring 
precession model \citep{Stella1998}, and the propagating fluctuation model \citep{Ingram2011}, 
etc. Using their propagating fluctuation model, \cite{Ingram2013} explained the 
anti-correlation between the short-term time lag and the Fourier frequency well. There 
may be a prospect of interpreting the short-term time lags observed in XRBs with the 
help of these QPO models, whereas it is out of the scope of this paper.

Certainly, Comptonization is a common physical process for producing X-rays in XRBs, so 
it is widely invoked to interpret the observed time lags in the order of milliseconds in 
these sources \citep{vanderKlis1987,Ford1999,Nobili2000,Lee2001,Poutanen2001}. In the 
Comptonization model, the low-energy seed photons, emitted from a relatively cool region 
such as the accretion disk, are inversely Comptonized by the energetic electrons from a 
hot source such as a hot corona or the hot plasma near the compact objects, in which the 
low-energy photons gain energy, while the high-energy electrons lose part of their energy, 
leading to that high-energy photons undergo more inverse Compton scatterings and, 
therefore, they escape later than the low-energy photons, resulting in hard 
X-ray time lags \citep{Ford1999}. Obviously, soft X-ray time lags cannot be explained by 
this model. \cite{vanderKlis1987} proposed that the softening of shots could account for 
soft time lags. In order to explain the time lags of GRS 1915+105, \cite{Nobili2000} 
proposed a scenario in which a standard thin disk \citep{Shakura1973} coexists with a 
central corona with two parts, an inner part with relatively high temperature and an outer 
part with relatively low temperature. In the inner part, the inverse Compton scattering is 
taken place, resulting in hard X-ray time lags. However, in the outer part, the disk seed 
photons are Comptonized by the relatively cool electrons, in which the electrons gain energy, 
while the seed photons lose energy, so the low-energy photons undergo more Compton 
scatterings and escape later than high-energy photons, leading to soft X-ray time lags. 
\cite{Qu2001} invoked this scenario to explain the detected soft and hard X-ray time lags 
less than ten milliseconds in Cir X-1. In addition to the two models, there are some other 
models for explaining the short-term time lags of XRBs, such as the magnetic flare model 
\citep{Poutanen1999}. In this model, it is assumed that the X-rays are produced in compact 
magnetic flares and the movement for magnetic loops to inflate and detach from the 
accretion disk induces spectral evolution, which results in time lags corresponding to 
the evolution timescales of the flares. 

\subsection{The long-term time lags in XRBs}

Through CCF technique, the long-term time lags ranging from hundreds of seconds to 
over one thousand seconds between hard and soft X-rays have been detected in a few 
Galactic BH X-ray binaries (BHXBs). Anti-correlated long-term hard X-ray time lags 
were first detected in the low/hard state of the high-mass BHXB Cyg X-3 
\citep{Choudhury2004}, then this kind of hard X-ray time lags were found in 
microquasar GRS 1915+105 during its $\chi$ states \citep{Choudhury2005}. It is noted 
that the $\chi$ states of this microquasar are the closest analog to the low/hard 
states of BHXBs. In addition, the similar long-term anti-correlated hard X-ray time 
lags were also detected in another microquasar XTE J1550-564, but in its very high 
state or steep power law state \citep{Sriram2007}. Interestingly, the long-term soft 
X-ray time lags of microquasar GX 339-4 were found in its hard and soft intermediate 
states \citep{Sriram2010}. It is noted that the spectral pivoting was shown during 
the episodes when this kind of time lags were detected in these BHXBs 
\citep{Choudhury2004,Choudhury2005,Sriram2007}. \cite{Choudhury2004} proposed a 
truncated accretion disk model to explain the detected anti-correlated long-term 
hard X-ray time lags in BHXBs. In the scenario of this model, the optically thick 
accretion disk, where the soft X-rays are emitted, is truncated far away from the 
BH, while the truncated area, i.e. the high-temperature region between the inner 
disk edge and the location near the BH, is full of Compton cloud which is 
responsible for the hard X-ray emission. Any change in the disk will trigger 
corresponding anti-correlated change in the Compton cloud in a viscous timescale, 
during which the accreting matter flows from the optically thick accretion disk to 
the truncated area, resulting in that the hard X-rays emitted in the Compton cloud 
lag behind the soft X-rays emitted in the disk.

Through the CCF method, the anti-correlated long-term hard time lags from tens of 
seconds to hundreds of seconds were also detected in a few NS-LMXBs. \cite{Lei2008} 
first detected anti-correlated long-term hard X-ray time lags in Z source Cyg X-2 
and, meanwhile, anti-correlated long-term soft X-ray time lags were also found in 
this source. \cite{Sriram2012} reported the similar anti-correlated long-term hard 
and soft X-ray time lags detected in another Z source GX 5-1. Moreover, in atoll 
source 4U 1735-44 and 4U 1608-52, \cite{Lei2013,Lei2014} detected four types of 
long-term X-ray time lags, i.e. anti-correlated long-term hard and soft X-ray time 
lags and positively correlated long-term hard and soft X-ray time lags. \cite{Wang2014} 
systematically investigated the cross-correlation evolution of XTE J1701-462 on its 
HIDs, detected the four types of long-term X-ray time lags, and found that its 
cross-correlation behavior evolves with luminosity. The model of truncated accretion 
disk was invoked to explain the detected long-term time lags in NS-LMXBs 
\citep{Lei2008,Sriram2012,Wang2014}. However, we argue that the model of truncated 
accretion disk cannot account for the detected long-term time lags in NS-LMXBs. Firstly, 
this model can only account for the anti-correlated long-term hard X-ray time lags, 
while anti-correlated long-term soft X-ray time lags are usually detected in NS-LMXBs. 
Secondly, the truncated accretion disk model requires that an optically thick 
accretion disk is far away from the compact object, while the accretion disk of 
NS-LMXBs touches the NS \citep{Inogamov1999,Inogamov2010}. Here, we argue that in 
NS-LMXBs, the accretion disk touches the NS because of the absence of a magnetoshpere 
around the NS in these systems. In the XRBs hosting a strongly magnetized NS with 
surface magnetic field strength ${\rm B_0\sim10^{12}\ G}$, e.g. accretion-powered 
X-ray binary pulsars, most of which are NS-HMXBs, a magnetosphere will be formed 
around the NS \citep{Lamb1973}. In these XRBs, when the accreting matters approach the 
magnetosphere, they will be channeled onto the polar caps of the NS along the magnetic 
field lines, producing X-ray pulsations, which could be an evidence for a magnetosphere 
in XRBs. Additionally, cyclotron lines are usually observed in the X-ray spectra of 
accretion X-ray pulsars \citep{Coburn2002,Caballero2012}, showing sufficient ${\rm B_0}$ 
to form a magnetosphere. Sequentially, in NS-HMXBs, the magnetic pressure from the 
magnetosphere pushes the accretion disk outwards, leading to that the disk might be far 
away from the NS. However, ${\rm B_0}$ of NS-LMXBs, including Z sources and atoll 
sources, spans a range of ${\rm \sim10^{7}-10^{9}\ G}$ \citep{Focke1996,Campana2000,
Ding2006,Ding2011}, which is much smaller than that of strongly magnetized NSs, so that 
it is unlikely for a magnetosphere to be formed in NS-LMXBs. Actually, the X-ray 
pulsations have never been observed in Z sources and atoll sources except Aql X-1 
\citep{Casella2008} and cyclotron lines have never been observed in these sources too, 
any of which conforms the absence of a magnetosphere in NS-LMXBs. In the case of the 
absence of a magnetosphere, the accretion disk will reach the NS. Therefore, the 
accretion disk of NS-LMXBs touches the NS.
    
\subsection{The long-term time lags in GX 349+2}

In this work, using all the 138 {\it RXTE} observations from 1996 to 2011 for Z 
source GX 349+2, we systematically perform cross-correlation analysis between 2-5 
kev and 16-30 keV light curves with CCF method. The results are listed in 
Tables~\ref{table1}-\ref{table3} and selectively shown by 
Figures~\ref{fig:positive}-\ref{fig:anti-correlation}.

As reviewed and pointed out above, the long-term hard X-ray time lags of BHXBs can 
be explained by the truncated accretion disk model 
\citep{Choudhury2004,Choudhury2005,Sriram2007}, but this model cannot account for 
the long-term time lags of NS-LMXBs, because the accretion disk in NS-LMXBs actually 
contacts with the NS \citep{Church2004,Inogamov1999,Inogamov2010}. Moreover, the 
long-term time lags detected in XRBs, including the results that we obtain in this 
work, are larger than dozens of seconds, even over one thousand seconds, while the 
short-term time lags in XRBs spans in the range of milliseconds, so it is a feasible 
assumption that the mechanisms responsible for the long-term X-ray time lags could 
be very different from those models accounting for the short-term time lags in XRBs, 
such as the shot model reviewed above. Besides, in general, the short-term time lags 
are derived from two light curves in two adjacent energy bands, e.g. 2-5 keV vs. 5-13 
keV for GRS 1915+105 \citep{Reig2000} and 1.8-5.1 keV vs. 5.1-13.1 keV for Cir X-1 
\citep{Qu2001}, while the long-term time lags are usually derived from two light 
curves with distant energy bands, e.g. 2-7 keV vs. 20-50 keV for Cyg X-3 
\citep{Choudhury2004} and 2-5 keV vs. 16-30 keV for GX 5-1 \citep{Sriram2012}, which 
indicates that the X-rays responsible for the short-term time lags come from two 
adjacent emission regions, even from the same area, while the the X-rays producing 
long-term time lags might come from two regions which are far away each other. 

Here, we invoke an extend accretion disk corona (ADC) model to explain the 
long-term time lags detected in NS-LMXBs, including the long-term time lags of 
GX 349+2 that we derive in this work. Analyzing the dip and non-dip spectra of 
NS-LMXBs, \cite{Church1993,Church1995} proposed a Birmingham model which consists 
of a blackbody component interpreted as the emission from a point source, i.e. the 
NS, and a power law component that might be resulted from the Comptonization of 
thermal emission in an ADC above the accretion disk. Through dip ingress time 
technique, \cite{Church2004} measured the radius of the ADC and developed the 
Birmingham model into an extended ADC model. The measured radius of a thin, hot 
corona above the accretion disk varies in the range of $\sim$(2--70)$\times10^4$~km. 
Therefore, the corona is very extended and the disk is substantially covered by the 
corona. In the extended ADC model, almost all soft X-ray photons from the accretion 
disk are inversely Comptonized by the energetic electrons from the extended ADC, which 
produces the observed hard X-rays, while the observed soft X-rays are interpreted 
as the emission from the NS; the accretion disk is illuminated by the emission of 
the NS, leading to the production of the extended ADC above the disk. This model 
was successfully applied to Cyg-like Z sources \citep{abc2006,Jackson2009,Balucinska2010} 
and Sco-like Z sources (including GX 349+2) \citep{Church2012}, as well as atoll 
sources \citep{Church2014}, so it could be a universal model for NS-LMXBs. Since 
the extended ADC model is a unified model for NS-LMXBs, we try to interpret the 
long-term time lags in NS-LMXBs with the help of this model. The extended ADC and 
the NS are two independent emitting regions, which satisfies the request that the 
hard and soft X-rays for long-term time lags are emitted from two distant regions, 
as discussed above. In order to explain the long-term time lags detected in NS-LMXBs 
in terms of the extended ADC model, we introduce two timescales. One is the 
Comptonization timescale during which the disk seed photons are inversely Comptonized 
by the high-energy electrons in the extended ADC, and another is the viscous timescale 
in the order of hundreds of seconds \citep{Lei2008}, during which the accreting matter 
flows from the disk onto the NS. The hard X-ray time lags will be produced if the 
Comptonization timescale is less than the viscous timescale, and, contrarily, the 
soft X-ray time lags will be observed if the Comptonization timescale is larger than 
the viscous timescale. It is noted that a minority of positively correlated short-term 
time lags ($<$1 s) are listed in Tables~\ref{table1} and \ref{table3}, which are derived 
with the CCF method in our work. These short-term time lags cannot be explained by the 
models reviewed in section 4.1, because those models are used to interpret the 
short-term time lags produced in two adjacent energy intervals, while these short-term 
time lags obtained in this work are derived from two distant energy intervals, i.e. 
2-5 kev and 16-30 keV energy intervals. In the frame of the extended ADC model, we 
propose that these short-term time lags will be observed under the circumstance that 
the Comptonization timescale is comparable with the viscous timescale. 

\section{CONCLUSION}

In this work, using all the {\it RXTE} observations for Z source GX 349+2, we 
systematically perform cross-correlation analysis between soft and hard X-rays with 
CCF method. Positive correlations and corresponding hard and soft X-ray long-term 
time lags are detected throughout an extensive Z track on its HID. Anti-correlated 
correlations and anti-correlated soft and hard X-ray long-term time lags are found 
outside the HID. It is noted that in most observations outside the HID, positive 
correlations and positively correlated hard and soft long-term X-ray time lags are 
also obtained. We review the short-term time lags obtained with the Fourier 
cross-spectral analysis in XRBs and some models responsible for the short-term time 
lags. We also review the long-term hard X-ray time lags found in BHXBs, which can 
be explained by the truncated accretion disk model. We argue that the long-term 
X-ray time lags in NS-LMXBs cannot be interpreted by those models responsible for 
the short-term time lags or the truncated accretion disk model. We invoke the 
extended ADC model to explain the long-term X-ray time lags in NS-LMXBs, including 
GX 349+2.  

\section*{Acknowledgements}

We thank the anonymous referee for her or his constructive comments and suggestions, 
which helped us carry out this research and improve the presentation of this paper. 
This research has made use of the data obtained through the High Energy Astrophysics 
Science Archive Research Center (HEASARC) On-line Service, provided by NASA/Goddard 
Space Flight Center (GSFC). This work is partially supported by the 2014 Project of 
Xinjiang Uygur Autonomous Region of China for Flexibly Fetching in Upscale Talents, 
National Key Basic Research Program of China (973 Program 2015CB857100), the Natural 
Science Foundation of China under grant nos. 11173024 and 11203064, and the Program 
of the Light in Chinese Western Region (LCWR) under grant no. XBBS 201121 provided 
by Chinese Academy of Sciences (CAS). 

\bibliographystyle{mn2e}

\clearpage

\begin{table*}
\vspace{6.5cm}
\footnotesize
\caption{The results of cross-correlation analysis on the HID.}\label{table1}
\vspace{-2.mm}
\begin{flushleft}
\begin{tabular}{lccccc}
\hline\hline
$^{a}$HID Region & $^{b}$Position & $^{c}$Live Time (s) & $^{d}$CCC & $^{e}$Time Lag (s) & $^{f}$Hardness \\
\hline
     1     & UNB & 2023 & 0.68$\pm$0.05 & 4.0$\pm$1.2   & 0.78   \\
     2     & UNB & 3936 & 0.71$\pm$0.13 & -1.0$\pm$3.0  & 0.76   \\
     3     & UNB & 4624 & 0.41$\pm$0.09 & 9.4$\pm$6.4   & 0.75   \\
     4     & MNB & 3752 & 0.72$\pm$0.13 & 2.0$\pm$2.6   & 0.73   \\
     5     & MNB & 3152 & 0.64$\pm$0.56 & 0.4$\pm$1.5   & 0.72   \\
     6     & MNB & 3648 & 0.68$\pm$0.08 & 0.7$\pm$2.1   & 0.70   \\
     9     & LNB & 6112 & 0.27$\pm$0.05 & -36.4$\pm$9.2 & 0.65   \\
    10     & LNB & 2432 & 0.53$\pm$1.07 & -6.1$\pm$7.5  & 0.64   \\
    11     & LFB & 8606 & 0.62$\pm$0.09 & -0.1$\pm$2.4  & 0.65   \\
    12     & LFB & 4960 & 0.74$\pm$0.09 & -2.6$\pm$1.9  & 0.66   \\
    13     & LFB & 3560 & 0.82$\pm$0.08 & -0.2$\pm$1.5  & 0.67   \\
    14     & LFB & 4032 & 0.53$\pm$0.07 &  1.3$\pm$3.7  & 0.68   \\
    15     & LFB & 4080 & 0.95$\pm$1.24 & -3.8$\pm$11.1 & 0.69   \\
    16     & MFB & 3360 & 0.88$\pm$0.88 & -3.6$\pm$8.1  & 0.70   \\
    17     & MFB & 3168 & 1.63$\pm$6.45 &  7.1$\pm$3.3  & 0.71   \\
    18     & MFB & 2224 & 0.79$\pm$0.14 &  0.1$\pm$2.9  & 0.72   \\
    22     & UFB & 1072 & 1.49$\pm$5.88 & -5.7$\pm$9.4  & 0.81   \\
    23     & UFB & 1232 & 0.87$\pm$0.11 &  0.8$\pm$2.2  & 0.83   \\
\hline
\vspace{-5.mm}
\end{tabular}
\end{flushleft}
\leftline{$^{a}$The regions in which positive correlations are detected.}
\leftline{$^{b}$The positions of the regions.}
\leftline{$^{c}$The total relative time length of each region.}
\leftline{$^{d}$The derived cross$-$correlation coefficients.}
\leftline{$^{e}$The derived time lags.}
\leftline{$^{f}$The hardness.}
\end{table*}

\clearpage

\begin{table*}
\vspace{7.cm}
\footnotesize
\caption{The results of anti-correlated correlations which are detected outside the period 
of the HID.}\label{table2}
\vspace{-2.mm}
\begin{flushleft}
\begin{tabular}{lccc}
\hline\hline
$^{a}$ObsID & $^{b}$Date & $^{c}$CCC & $^{d}$Time Lag (s)  \\
\hline
30043-01-07-00     & 1998-10-09  & -0.34$\pm$0.03 & 192$\pm$10\\
50017-01-02-00     & 2003-01-09  & -0.21$\pm$0.01 & 1177$\pm$57\\
80105-05-01-00     & 2003-09-23  & -0.72$\pm$0.02 & -6107$\pm$46\\
90024-04-02-00     & 2004-04-17  & -0.36$\pm$0.07 & -77$\pm$5 \\
90024-04-09-00     & 2004-08-02  & -0.24$\pm$0.03 & -327$\pm$16\\
90024-04-12-00     & 2004-09-22  & -0.33$\pm$0.04 & -66$\pm$10\\
90024-04-16-00     & 2005-01-08  & -0.28$\pm$0.06 & 85$\pm$6\\
90024-04-21-00     & 2005-04-02  & -0.24$\pm$0.05 & 258$\pm$20\\
93071-05-01-00     & 2008-09-17  & -0.33$\pm$0.02 & -127$\pm$15\\
93071-05-03-00     & 2008-09-26  & -0.16$\pm$0.01 & -25$\pm$27\\
\hline
\vspace{-5.mm}
\end{tabular}
\end{flushleft}
\leftline{$^{a}$The observations in which anti-correlated correlations are detected.}
\leftline{$^{b}$The observation dates.}
\leftline{$^{c}$The derived cross$-$correlation coefficients.}
\leftline{$^{d}$The derived time lags.}
\end{table*}

\clearpage

\begin{table*}
\vspace{1.cm}
\footnotesize
\caption{The results of positive correlations which are detected outside 
the period of the HID.}\label{table3}
\vspace{-2.mm}
\renewcommand{\arraystretch}{1.05}
\renewcommand{\tabcolsep}{1.0pc}
\begin{flushleft}
\begin{tabular}{lccc}
\hline \hline
$^{a}$ObsID & $^{b}$Date & $^{c}$CCC & $^{d}$Time Lag (s)  \\
\hline
10063-11-01-00     & 1996-09-07  &  0.62$\pm$0.04 & 4.8$\pm$5.1\\
10063-12-01-00     & 1996-09-06  &  0.69$\pm$0.01 & 7.6$\pm$23.2\\
30043-01-01-00     & 1998-09-29  &  0.50$\pm$0.03 & 1.1$\pm$46\\
30043-01-02-00     & 1998-09-29  &  0.89$\pm$0.01 & -2.5$\pm$2.9 \\
30043-01-03-00     & 1998-09-30  &  0.35$\pm$0.03 & 39.3$\pm$9.6\\
30043-01-04-00     & 1998-09-30  &  0.85$\pm$0.03 & 24.1$\pm$7.7\\
30043-01-05-00     & 1998-10-01  &  0.78$\pm$0.02 & -29.5$\pm$5.7\\
30043-01-06-00     & 1998-10-01  &  0.85$\pm$0.03 & 8.6$\pm$2.9\\
30043-01-08-00     & 1998-10-02  &  0.91$\pm$0.02 & 6.8$\pm$4.5\\
30043-01-09-00     & 1998-10-03  &  0.77$\pm$0.03 & -18.7$\pm$5.9\\
30043-01-10-00     & 1998-10-03  &  0.54$\pm$0.03 & 6.3$\pm$7.8\\
30043-01-11-00     & 1998-10-04  &  0.65$\pm$0.03 & -18.0$\pm$14.1 \\
30043-01-12-00     & 1998-10-04  &  0.84$\pm$0.02 & 0.8$\pm$3.9\\
30043-01-13-00     & 1998-10-05  &  0.53$\pm$0.06 & 2.2$\pm$4.8\\
30043-01-14-00     & 1998-10-05  &  0.64$\pm$0.02 & -18.1$\pm$6.0\\
30043-01-15-00     & 1998-10-06  &  0.63$\pm$0.04 & 64.4$\pm$14.0\\
30043-01-16-00     & 1996-09-07  &  0.94$\pm$0.05 & -18.7$\pm$16.7\\
30043-01-17-00     & 1998-10-07  &  0.54$\pm$0.03 & 38.1$\pm$5.9\\
30043-01-18-00     & 1998-10-08  &  0.69$\pm$0.06 & -18.1$\pm$14.4\\
30043-01-19-00     & 1998-10-08  &  0.80$\pm$0.03 & -3.4$\pm$6.7 \\
30043-01-20-00     & 1998-10-09  &  0.51$\pm$0.06 & -4.0$\pm$7.7\\
30043-01-21-00     & 1998-10-09  &  0.82$\pm$0.03 & -7.8$\pm$5.1\\
30043-01-22-00     & 1998-10-10  &  0.84$\pm$0.05 & -29.3$\pm$16.0\\
30043-01-23-00     & 1998-10-10  &  0.83$\pm$0.07 & -3.6$\pm$4.73\\
30043-01-24-00     & 1998-10-11  &  0.88$\pm$0.03 & -13.0$\pm$9.7\\
30043-01-25-00     & 1998-10-11  &  0.33$\pm$0.05 & 7.7$\pm$12.0\\
30043-01-26-00     & 1998-10-12  &  0.54$\pm$0.04 & 5.1$\pm$17.0\\
30043-01-27-00     & 1998-10-12  &  0.70$\pm$0.05 & -5.4$\pm$10.8 \\
30043-01-28-00     & 1998-10-13  &  0.86$\pm$0.05 & 0.6$\pm$5.8\\
30043-01-29-00     & 1998-10-13  &  0.42$\pm$0.06 & 2.2$\pm$4.8\\
30043-01-30-00     & 1998-10-14  &  0.32$\pm$0.05 & -30.7$\pm$10.3\\
50017-01-01-00     & 2000-03-27  &  0.20$\pm$0.04 & 42.5$\pm$20.2\\
50017-01-01-02     & 2000-03-29  &  0.85$\pm$0.05 & -2.6$\pm$4.0\\
50017-01-01-03     & 2000-03-29  &  0.78$\pm$0.02 & -20.6$\pm$8.1\\
50017-01-01-04     & 2000-03-29  &  0.74$\pm$0.02 & 3.7$\pm$7.9\\
50017-01-01-06     & 2000-03-29  &  0.68$\pm$0.05 & -16.2$\pm$8.3 \\
50017-01-01-08     & 2000-03-30  &  0.57$\pm$0.06 & 6.7$\pm$4.2\\
50017-01-01-09     & 2000-03-31  &  0.62$\pm$0.05 & -3.1$\pm$12.4\\
90022-03-01-00     & 2004-08-05  &  0.82$\pm$0.05 & 1.5$\pm$8.9\\
90022-03-02-00     & 2004-08-21  &  0.77$\pm$0.03 & -6.8$\pm$9.2\\
90022-03-03-00     & 2004-08-28  &  0.73$\pm$0.01 & -13.6$\pm$5.9\\
90022-03-04-00     & 2004-09-04  &  0.77$\pm$0.07 & 10.2$\pm$4.6\\
90022-03-04-01     & 2004-09-05  &  0.86$\pm$0.02 & 47.7$\pm$19.9\\
90022-03-04-02     & 2004-09-07  &  0.25$\pm$0.02 & -13.6$\pm$10.1 \\
90022-03-04-03     & 2004-09-07  &  0.37$\pm$0.05 & -29.3$\pm$11.9\\
90022-03-05-00     & 2004-09-15  &  0.85$\pm$0.05 & -26.6$\pm$12.6\\
90022-03-05-01     & 2004-09-14  &  0.81$\pm$0.03 & 9.5$\pm$7.4\\
90022-03-07-00     & 2005-10-05  &  0.61$\pm$0.07 & -13.3$\pm$10.8\\
90022-03-07-01     & 2005-10-06  &  0.84$\pm$0.05 & -8.2$\pm$15.1\\
90022-03-08-00     & 2005-10-07  &  0.82$\pm$0.06 & -0.2$\pm$6.2\\
90022-03-08-01     & 2005-10-07  &  0.69$\pm$0.02 & 1.4$\pm$6.2\\
90022-03-09-00     & 2006-10-12  &  0.32$\pm$0.03 & 30.4$\pm$14.2 \\
90024-04-01-00     & 2004-04-02  &  0.75$\pm$0.05 & -4.1$\pm$4.2\\
\end{tabular}
\end{flushleft}
\end{table*}

\clearpage

\begin{table*}
\vspace{1.cm}
\contcaption{}
\renewcommand{\arraystretch}{1.05}
\renewcommand{\tabcolsep}{1.0pc}
\begin{flushleft}
\begin{tabular}{lccc}
\hline
$^{a}$ObsID & $^{b}$Date & $^{c}$CCC & $^{d}$Time Lag (s)  \\
\hline
90024-04-02-01     & 2004-04-18  &  0.96$\pm$0.11 & 0.3$\pm$6.3\\
90024-04-02-02     & 2004-04-22  &  0.91$\pm$0.17 & -1.9$\pm$8.4\\
90024-04-03-00     & 2004-05-21  &  0.54$\pm$0.04 & 20.8$\pm$16.0\\
90024-04-04-00     & 2004-05-28  &  0.80$\pm$0.07 & -0.6$\pm$5.2\\
90024-04-06-00     & 2004-06-14  &  0.85$\pm$0.06 & 6.6$\pm$8.6\\
90024-04-07-00     & 2004-07-01  &  0.37$\pm$0.06 & 102.8$\pm$14.6\\
90024-04-08-00     & 2004-07-16  &  0.75$\pm$0.07 & -0.1$\pm$3.2 \\
90024-04-10-00     & 2004-08-13  &  0.84$\pm$0.10 & -3.9$\pm$3.2\\
90024-04-11-00     & 2004-09-09  &  0.27$\pm$0.11 & -13.8$\pm$8.4\\
90024-04-13-00     & 2004-10-12  &  0.74$\pm$0.05 & 5.2$\pm$11.1\\
90024-04-14-00     & 2004-10-26  &  0.71$\pm$0.05 & -26.1$\pm$9.1\\
90024-04-15-00     & 2004-11-13  &  0.53$\pm$0.05 & 1.5$\pm$6.7\\
90024-04-17-00     & 2005-01-26  &  0.67$\pm$0.08 & 25.1$\pm$11.0\\
90024-04-18-00     & 2005-02-15  &  0.69$\pm$0.06 & 18.8$\pm$10.4\\
90024-04-18-01     & 2005-02-12  &  0.79$\pm$0.08 & -0.9$\pm$1.6\\
90024-04-18-02     & 2005-02-16  &  0.87$\pm$0.06 & -23.7$\pm$7.9\\
90024-04-19-00     & 2005-02-24  &  0.81$\pm$0.09 & 5.6$\pm$6.8\\
90024-04-20-00     & 2005-03-16  &  0.42$\pm$0.10 & -22.1$\pm$9.7 \\
90024-04-22-00     & 2005-05-17  &  0.82$\pm$0.09 & -13.7$\pm$8.3\\
90024-04-23-00     & 2005-05-25  &  0.87$\pm$0.10 & -4.0$\pm$3.2\\
90024-04-25-00     & 2005-06-17  &  0.80$\pm$0.05 & -48.4$\pm$11.8\\
90024-04-26-00     & 2005-07-11  &  0.24$\pm$0.05 & 1.6$\pm$10.3\\
90024-04-27-00     & 2005-07-19  &  0.21$\pm$0.04 & -4.0$\pm$6.1\\
90024-04-28-00     & 2005-08-11  &  0.90$\pm$0.09 & 0.2$\pm$1.6\\
90024-04-29-00     & 2005-08-25  &  0.38$\pm$0.03 & -49.1$\pm$10.4\\
90024-04-31-00     & 2005-09-21  &  0.76$\pm$0.07 & -5.1$\pm$2.3\\
90024-04-32-00     & 2005-10-10  &  0.60$\pm$0.08 & -0.8$\pm$2.4\\
90024-04-33-00     & 2005-10-17  &  0.37$\pm$0.04 & 15.0$\pm$11.7\\
90024-04-34-00     & 2005-11-02  &  0.90$\pm$0.05 & -42.7$\pm$9.9 \\
90024-04-35-00     & 2005-11-11  &  0.88$\pm$0.06 & 6.7$\pm$3.3\\
90024-04-36-00     & 2006-01-16  &  0.53$\pm$0.04 & -32.7$\pm$6.3\\
90024-04-37-00     & 2006-02-13  &  0.91$\pm$0.14 & -7.2$\pm$8.8\\
90024-04-38-00     & 2006-02-18  &  0.80$\pm$0.09 & 2.8$\pm$1.8\\
90024-04-39-00     & 2006-03-10  &  0.61$\pm$0.03 & 28.1$\pm$9.1\\
90024-04-42-00     & 2006-05-05  &  0.70$\pm$0.06 & 6.3$\pm$6.7\\
90024-04-43-00     & 2006-05-21  &  0.89$\pm$0.05 & 0.3$\pm$0.9\\
90024-04-44-00     & 2006-06-18  &  0.73$\pm$0.07 & -11.4$\pm$3.8\\
90024-04-45-00     & 2006-07-17  &  0.65$\pm$0.05 & -13.7$\pm$11.4\\
90024-04-46-00     & 2006-08-05  &  0.83$\pm$0.05 & 2.6$\pm$3.3\\
91152-01-01-00     & 2006-05-10  &  0.70$\pm$0.01 & 6.3$\pm$8.5 \\
91152-01-01-01     & 2006-05-10  &  0.87$\pm$0.03 & 3.7$\pm$6.4\\
91152-01-02-00     & 2006-07-04  &  0.85$\pm$0.02 & 135.7$\pm$27.4\\
91152-01-03-00     & 2006-08-20  &  0.62$\pm$0.02 & 16.0$\pm$18.9\\
93406-07-01-00     & 2007-07-05  &  0.81$\pm$0.03 & 50.0$\pm$13.5\\
96378-04-01-00     & 2011-07-06  &  0.46$\pm$0.01 & -286.7$\pm$40.1\\
96378-04-01-01     & 2011-07-06  &  0.87$\pm$0.05 & -1.4$\pm$9.4\\
96378-04-02-00     & 2011-10-09  &  0.74$\pm$0.01 & -76.8$\pm$19.6\\
\hline
\vspace{-5.mm}
\end{tabular}
\end{flushleft}
\leftline{$^{a}$The observations in which positive correlations are detected.}
\leftline{$^{b}$The observation dates.}
\leftline{$^{c}$The derived cross$-$correlation coefficients.}
\leftline{$^{d}$The derived time lags.}
\end{table*}

\clearpage

\clearpage

\clearpage

\clearpage

\clearpage

\end{document}